\pdfoutput=1

\documentclass[11pt,twoside,a4paper,cmspaper,final,collab]{cms-tdr}
\def\svnVersion{467129P}\def\cmsCernNoTag{CERN-EP-2017-236}\def\cmsCernDate{\today}\def\cmsMessage{Published in the Journal of High Energy Physics as \href{http://dx.doi.org/10.1007/JHEP04(2018)033}{\DOI{10.1007/JHEP04(2018)033}}}

\begin{document}\cmsNoteHeader{TOP-15-016}

\hyphenation{had-ron-i-za-tion}
\hyphenation{cal-or-i-me-ter}
\hyphenation{de-vices}
\RCS$Revision: 460337 $
\RCS$HeadURL: svn+ssh://svn.cern.ch/reps/tdr2/papers/TOP-15-016/trunk/TOP-15-016.tex $
\RCS$Id: TOP-15-016.tex 460337 2018-05-15 11:41:48Z chwalek $
\renewcommand{\cmsPubBlock}{\begin{tabular}[t]{@{}r@{}l} &CMS \cmsSTYLE\ \cmsNUMBER\\ &ATLAS ATL-XXX\\ \end{tabular}}
\renewcommand{\cmslogo}{\includegraphics[height=2cm]{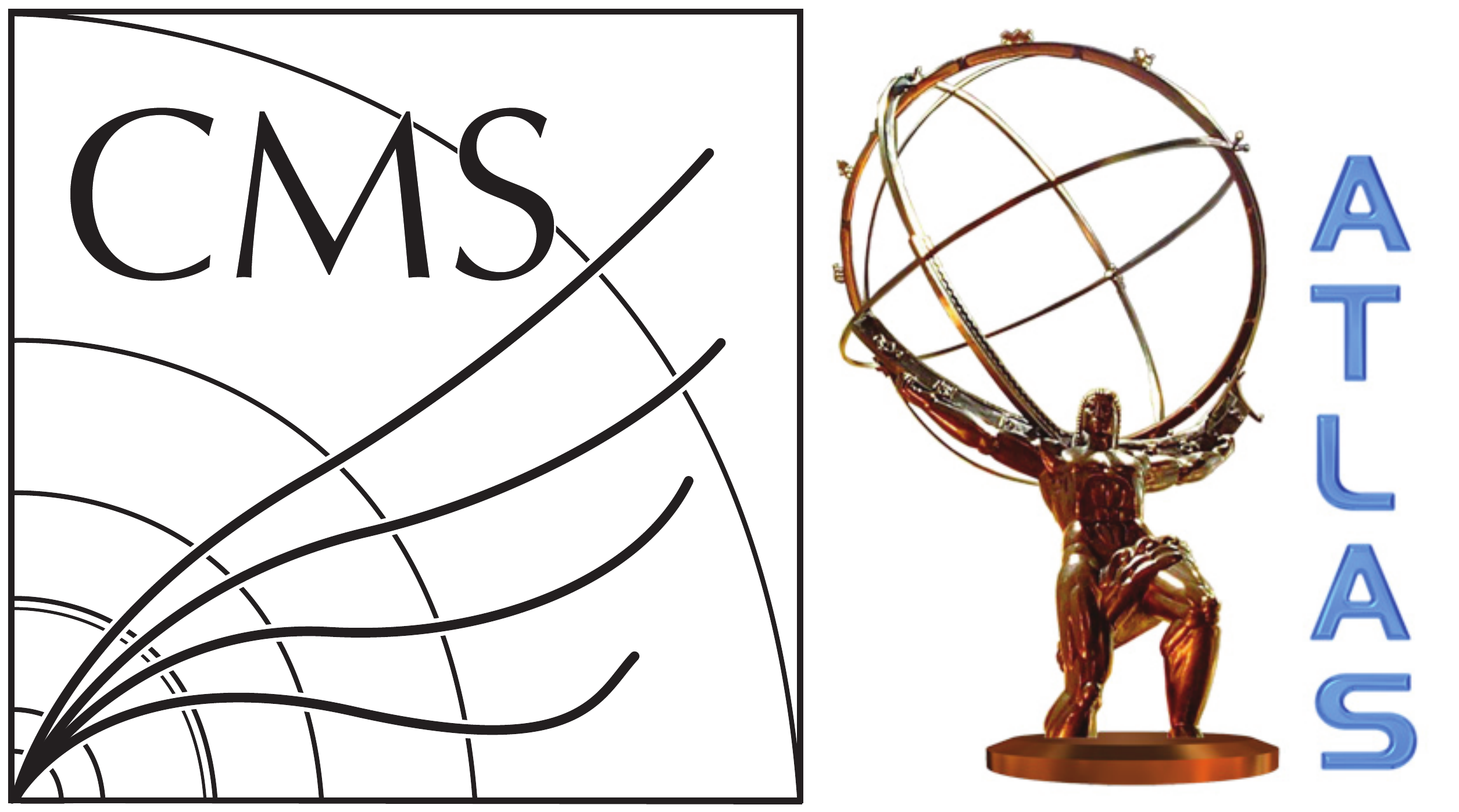}}
\renewcommand{\cmsCollabName}{The ATLAS and CMS Collaborations}
\renewcommand{\cmsCernNoTag}{\relax}
\renewcommand{\cmsTag}{\begin{tabular}[t]{@{}r@{}l} &CMS  \cmsNUMBER\\ &ATLAS-TOPQ-2016-16\\ \end{tabular}}
\renewcommand{\appMsg}{See Appendices~\ref{app:collab} and \ref{app:ATLAScollab} for the lists of collaboration members}
\providecommand{\ptmiss}{\ensuremath{p_{\mathrm{T}}^{\text{miss}}}\xspace}
\providecommand{\corr}{\ensuremath{\mathrm{corr}}}

\cmsNoteHeader{TOP-15-016}
\title{Combination of inclusive and differential \ttbar charge asymmetry measurements
using ATLAS and CMS data at $\sqrt{s} = 7$ and 8\TeV}

\date{\today}

\abstract{This paper presents combinations of inclusive and differential
  measurements of the charge asymmetry ($A_{\mathrm{C}}$) in top quark pair ($\ttbar$) events
  with a lepton+jets signature by the ATLAS and CMS Collaborations, using data
  from LHC proton-proton collisions at centre-of-mass energies of
  7 and 8\TeV. The data correspond to integrated
  luminosities of about 5 and $20\fbinv$ for
  each experiment, respectively. The resulting combined LHC measurements of the inclusive charge asymmetry are $A_{\mathrm{C}}^{\mathrm{LHC7}} = 0.005 \pm0.007\stat\pm0.006\syst$ at 7\TeV and $A_{\mathrm{C}}^{\mathrm{LHC8}} = 0.0055 \pm0.0023\stat\pm0.0025\syst$ at 8\TeV. These values, as well as the combination of $A_{\mathrm{C}}$ measurements as a function of the invariant mass of the $\ttbar$ system at 8\TeV, are  consistent with the respective standard model predictions.}

\hypersetup{%
pdfauthor={CMS Collaboration},%
pdftitle={Combination of inclusive and differential ttbar charge asymmetry measurements
using ATLAS and CMS data at 7 and 8 TeV},%
pdfsubject={CMS},%
pdfkeywords={CMS, ATLAS, physics, top quark, asymmetry, combination, LHCtopWG}}

\maketitle

\section{Introduction}

With the large number of top quark pair ($\ttbar$) events produced at the CERN LHC, the properties of the most massive elementary particle known to date are studied with ever increasing accuracy. Asymmetries in the angular distributions of top quarks and antiquarks provide powerful probes for physics beyond the standard model (SM). In proton-antiproton collisions at the Tevatron, it is possible to define a forward-backward asymmetry $A_{\mathrm{FB}}$~\cite{CDF,Dzero, Aaltonen:2017efp},
while the same underlying physical effects induce a charge asymmetry $A_{\mathrm{C}}$ in proton-proton ($\Pp\Pp$) collisions at the LHC~\cite{Kuhn,Aguilar-Saavedra:2014kpa}.

The production of $\ttbar$ pairs via gluon fusion is symmetric with respect to the exchange of the top quark and antiquark. The same is true for the $\qqbar\to \ttbar$ process at leading order (LO) in quantum chromodynamics (QCD). Asymmetries in kinematic observables of top quarks and antiquarks are introduced by higher-order effects in QCD for events produced by quark-antiquark annihilation. Interference effects connect the direction of motion of the top quark to that of the incoming quark, and the direction of motion of the top antiquark to that of the incoming antiquark~\cite{Kuhn:1998jr}. Initial quark and antiquark momenta in the protons have different spectra, leading to a measurable difference between the angular distributions of top quark and antiquark in $\Pp\Pp$ collisions. On average, quarks (valence and sea quarks) carry larger momentum than the antiquarks (sea quarks), causing the rapidity distribution of top quarks to be broader than that of top antiquarks.

In proton-proton collisions, the $\ttbar$ charge asymmetry is defined as
\begin{linenomath}
\begin{equation}
A_{\mathrm{C}} = \frac{N^{\Delta\abs{y} > 0 } - N^{\Delta\abs{y} < 0}}{N^{\Delta\abs{y} > 0}+N^{\Delta\abs{y} < 0}},
\end{equation}
\label{eq:Ac}
\end{linenomath}
using the difference of the absolute values of the rapidities $y$ of the top quark and antiquark, $\Delta\abs{y} = \abs{y_{\PQt}} - \abs{y_{{\PAQt}}}$, as the sensitive observable. The numbers of events with $\Delta\abs{y}$ taking positive or negative values are given by $N^{\Delta\abs{y} > 0}$ and $N^{\Delta\abs{y} < 0}$, respectively.

The ATLAS~\cite{Aad:2008zzm} and CMS~\cite{Chatrchyan:2008aa} Collaborations have measured the inclusive $\ttbar$ charge asymmetry at centre-of-mass energies of 7 and 8\TeV in events with one charged lepton (lepton+jets channel)~\cite{ATLAS_7TeV, CMS_7TeV, ATLAS_8TeV,CMS_8TeV_Template, CMS_8TeV_Unfolding} and in events with two charged leptons (dilepton channel) in the final state~\cite{Aad:2015jfa,Chatrchyan:2014yta,Aad:2016ove,Khachatryan:2016ysn}, where the leptons are either electrons or muons. The ATLAS Collaboration has also measured the charge asymmetry in highly boosted $\ttbar$ events, where the asymmetry is predicted to be amplified~\cite{Aad:2015lgx}. In addition, both collaborations have measured the charge asymmetry differentially as a function of suitable kinematic variables.

Theoretical predictions from QCD calculations are available at next-to-leading-order (NLO) from Refs.~\cite{Kuhn} and \cite{Bernreuther} and at next-to-next-to-leading-order (NNLO)~\cite{nnlo2, nnlo3, nnlo4, nnlo1} precision in the strong coupling. These calculations include electroweak (EW) corrections at NLO precision. The two calculations at NLO follow the same approach to evaluate the charge asymmetry but differ in technical details. The calculation in Ref.~\cite{Kuhn} uses a LO parton distribution function (PDF) set to evaluate the asymmetry, while the calculation in Ref.~\cite{Bernreuther} uses a NLO PDF set. The factorisation and renormalisation scales are set to the partonic centre-of-mass energy in Ref.~\cite{Kuhn}, while the calculation in Ref.~\cite{Bernreuther} uses fixed scales and sets both quantities to the top quark mass. The NNLO prediction is based on the methods described in Refs.~\cite{nnlo2, nnlo4}, derived using dynamical factorisation and renormalisation scales~\cite{nnlo3} ($\mu =H_{\mathrm{T}}/4$, where $H_{\mathrm{T}}=\sqrt{\smash[b]{m^2_{\PQt}+p^2_{\mathrm{T},\PQt}} }+ \sqrt{\smash[b]{m^2_{\PQt}+p^2_{\mathrm{T},\PAQt}}}$, with $m_{\PQt}$ being the top quark mass and $p_{\mathrm{T},\PQt/\PAQt}$ being the transverse momentum of the top quark or antiquark)
and a NNLO PDF set. In the NLO calculations, the ratio in Eq.~(\ref{eq:Ac}) is evaluated in powers of the considered couplings (strong and electroweak), taking NLO corrections into account only in the numerator, while the denominator is evaluated with the LO matrix element. For the NNLO prediction, numerator and denominator are calculated at full QCD NNLO precision (and NLO for electroweak corrections) without any expansion in powers of the considered couplings. Although the asymmetry predicted in the SM is small, contributions from beyond the standard model (BSM) could alter its value, especially at high values of the invariant mass of the $\ttbar$ system.

This paper reports the results of combinations of the inclusive $A_{\mathrm{C}}$ measurements from the ATLAS and CMS Collaborations at $\sqrt{s} = 7$ and 8\TeV, and the combination of the ATLAS and CMS differential asymmetry measurements as a function of the invariant mass of the $\ttbar$ system at 8\TeV. For the three combinations, only the results in the lepton+jets channel are considered. Including in addition
the results in the dilepton channel does not lead to a gain in precision because of their large statistical uncertainty.
All these measurements are extrapolated to the full phase space of $\ttbar$ production.
Table~\ref{tab:Summary7and8} gives an overview of the recent predictions from theory for both centre-of-mass energies and of the experimental results that serve as input to the combinations described in this paper.

The paper is structured as follows: Section~\ref{sec:input} and Section~\ref{sec:blue} briefly describe the analyses whose results are used for the combination and the method used to combine the results; Section~\ref{sec:syst} lists the uncertainties and the assumed correlations between the measurements; results and studies on the stability of the combinations are presented in Sections~\ref{sec:results} and \ref{sec:stability}.

\begin{table}[t!]
\topcaption{Overview of the most recent theoretical predictions for the inclusive $A_{\mathrm{C}}$ at the LHC at $\sqrt{s} = 7$ and 8\TeV, along with the experimental results from the ATLAS and CMS Collaborations. The uncertainties in all theoretical predictions are dominated by the uncertainties due to scale variations. The uncertainties in the experimental CMS results are given separately as statistical (first contribution) and systematic (second contribution) uncertainties, while for the ATLAS results the total uncertainty is quoted.
The ATLAS and CMS experimental results are described in Section~\ref{sec:input}.}
\centering
  \resizebox{\textwidth}{!}{
\begin{tabular}{llll}
\hline
\multicolumn{2}{c}{Inclusive $A_{\mathrm{C}}$~~~~~~~~~~~~~~~~~~~~~~~~}  & \multicolumn{2}{r}{Centre-of-mass energy~~~~~~~~~~~~~~~~~~~~~~~~~~~~~~~~~~~}\\
   &                                           &~~~~~~ 7\TeV   &~~~~~~ 8\TeV   \\
\hline
Theoretical               & QCD NLO + EW NLO~\cite{Kuhn}        & $0.0115 \pm 0.0006$ & $0.0102 \pm 0.0005$ \\
predictions          & QCD NLO + EW NLO~\cite{Bernreuther} & $0.0123 \pm 0.0005$ & $0.0111 \pm 0.0004$ \\
                     & QCD NNLO + EW NLO~\cite{nnlo1}       &                  & $0.0095^{+0.0005}_{-0.0007}$  \\
\hline
Experimental         & ATLAS~\cite{ATLAS_7TeV,ATLAS_8TeV}          & ~~$0.006 \pm 0.010$   & $0.0090 \pm 0.0051$ \\
results              & CMS unfolding~\cite{CMS_7TeV,CMS_8TeV_Unfolding}    & ~~$0.004 \pm 0.010 \pm 0.011$ & $0.0010 \pm 0.0068 \pm 0.0037$ \\
                     & CMS template~\cite{CMS_8TeV_Template}      &                 & $0.0033 \pm 0.0026 \pm 0.0033$\\
\hline
\end{tabular}}

\label{tab:Summary7and8}
\end{table}

\section{Input measurements}
\label{sec:input}

\subsection{The ATLAS measurements}

The ATLAS results used for the combinations are based on data recorded at $\sqrt{s} = 7\TeV$~\cite{ATLAS_7TeV} and 8\TeV~\cite{ATLAS_8TeV}, corresponding to integrated luminosities of 4.7 and $20.3\fbinv$, respectively.

Very similar analysis strategies are used for the two centre-of-mass energies. The analysis of the 8\TeV data profits from a larger number of selected events and also samples of simulated events generated using higher-order Monte Carlo (MC) event generators than were available for Ref.~\cite{ATLAS_7TeV}. In the 7\TeV analysis, the $\ttbar$ signal events are simulated using the LO multi-parton matrix element event generator \textsc{alpgen} 2.13~\cite{alpgen} with the LO PDF set CTEQ6L1~\cite{Pumplin:2002vw}.
Background events from the production of single-top-quarks are simulated using the LO event generator \textsc{acermc} 3.8~\cite{Kersevan:2004yg} ($t$ channel) or the NLO event generator \textsc{mc$@$nlo} 4.01~\cite{Frixione:2002ik, Frixione:2008yi, Frixione:2005vw} ($\PQt\PW$ and $s$ channel). In the 8\TeV  analysis, the $\ttbar$ signal events~\cite{Frixione:2007nw} and single-top-quark background events~\cite{Re:2010bp, Alioli:2009je} are simulated using the NLO event generator \textsc{powheg-Box} 1.0~\cite{powheg, powheg2, powheg3}
using the CT10~\cite{CT10} PDF set. At both centre-of-mass energies, simulated events are produced assuming a top quark mass of $172.5\GeV$.
Production of a $\PW$ or $\Z$ boson in association with jets, hereafter referred to as $\PW$+jets and $\Z$+jets, respectively, is simulated using \textsc{alpgen} with the CTEQ6L1 PDF set.
For the 7\TeV analysis, the dominant  $\PW$+jets background is
normalised using control samples in data, based on the fact that the
production rate of $\PW^+$+jets is larger than that of $\PW^-$+jets. The flavour composition of this background component is also adjusted from data.
For the 8\TeV analysis, the $\PW$+jets process is normalised \textit{in situ} using data while extracting the asymmetry. Parton showering and the underlying event are modelled in all samples using \textsc{herwig} 6.5.20~\cite{Corcella:2000bw} and \textsc{jimmy} 4.31~\cite{Butterworth:1996zw} with a set of tuned parameters called the AUET2 tune~\cite{ATLAS:2011gmi} for the 7\TeV analysis, and \textsc{pythia}~6.4.27~\cite{pythia} with the Perugia2011C tune~\cite{Skands:2010ak} for the 8\TeV analysis. The multijet background normalisation and shape are estimated from data using the matrix method~\cite{MM}.

The final state of $\ttbar$ events in the lepton+jets channel features an electron or muon, a neutrino, two $\PQb$-quarks from the two top quark decays, and two light quarks from the hadronically decaying  $\PW$ boson. Therefore, events with exactly one high-$\pt$ isolated electron or muon candidate and at least four jets are selected. Jets that originated from the hadronisation of a $\PQb$-quark are identified via a multivariate algorithm~\cite{btag1, btag2}. The operation point of the algorithm used for this measurement corresponds to $70\%$ efficiency to tag $\PQb$-quark jets with a rejection factor of
about 130 for light-quark or gluon jets (the rejection factor is equal to the inverse of the probability to erroneously tag jets from light quarks or gluons). For the analysis at 7\TeV, at least one jet is required to be $\PQb$-tagged, while events without $\PQb$-tagged jets are kept in the 8\TeV analysis for the \textit{in situ} calibration of the $\PW$+jets background.

Events from multijet and $\Z$+jets production rarely feature neutrinos and thus are expected to have only small missing transverse momentum ($\ptmiss$). The missing transverse momentum vector $\ptvecmiss$ is defined as the projection onto the plane perpendicular to the beam axis of the negative vector sum of all reconstructed objects in an event, and $\ptmiss$ is its magnitude. In order to suppress these backgrounds, requirements are imposed on the $\ptmiss$ value and the highly correlated transverse mass ($m_{\mathrm{T}}^{\PW}$) of the $\PW$ boson candidate from the semi-leptonically decaying top quark, formed by the lepton and $\ptvecmiss$. In the 7\TeV analysis, the \ptmiss value is required to be larger than $30~(20)\GeV$  in the electron+jets (muon+jets) channel. In addition, the $m_{\mathrm{T}}^{\PW}$ value is required to be larger than $30\GeV$ in the electron+jets channel while the sum of \ptmiss and $m_{\mathrm{T}}^{\PW}$ must be larger than $60\GeV$ in the muon+jets channel. For the 8\TeV analysis, the requirements are slightly different: $m_{\mathrm{T}}^{\PW} + \pt^{\text{miss}} > 60\GeV$ for  events with exactly zero or one $\PQb$-tagged jet, and $\pt^{\text{miss}} > 40~(20)\GeV$ for events with exactly zero (one) $\PQb$-tagged jet. In events with two or more $\PQb$-tagged jets, no requirements on $\ptmiss$ or $m_{\mathrm{T}}^{\PW}$ are made.

To measure the charge asymmetry, the full $\ttbar$ system is reconstructed and the $\Delta\abs{y}$ distribution is unfolded to parton level.
The reconstruction of the $\ttbar$ system is achieved by using a kinematic fit~\cite{klfitter} that assesses the compatibility of the observed event with the kinematic properties of $\ttbar$ events using a likelihood approach. In this approach, the most likely combination out of all
possible jet permutations is chosen. The observed $\Delta\abs{y}$ spectrum is unfolded to correct for acceptance and detector resolution effects using the fully Bayesian unfolding technique~\cite{fbu}. This method estimates the posterior probability density in bins of $\Delta\abs{y}$, which is integrated to define the asymmetry in Eq.~(\ref{eq:Ac}).
The treatment of systematic uncertainties is included by extending the likelihood entering the $\Delta\abs{y}$ posterior probability density computation with nuisance parameter terms that are marginalised.
In the 7\TeV analysis, the charge asymmetry is computed from the unfolded $\Delta\abs{y}$ distribution. At 8\TeV, a fit that maximises the extended likelihood over six event categories, defined based on the lepton charge and the $\PQb$-tagged jet multiplicity (zero $\PQb$-tagged jets, one $\PQb$-tagged jet, at least two $\PQb$-tagged jets), is performed. The $\PW$+jets calibration factors are fitted during the posterior probability estimation, as the $\PQb$-tagged jet multiplicity provides information about the heavy- and light-flavour composition of the $\PW$+jets background, while the lepton charge asymmetry is used to determine the normalisation of each $\PW$+jets component also during the posterior probability estimation. The signal fraction after applying the full selection is estimated to be 78\% for both centre-of-mass energies.

The inclusive $\ttbar$ charge asymmetries measured at 7 and 8\TeV can be found in Table~\ref{tab:Summary7and8}. The numbers quoted in Table~\ref{tab:Summary7and8} cannot be used directly for the combination, as a breakdown into correlated and uncorrelated components of the total uncertainty is needed. For the 7\TeV analysis, the result before marginalisation is used;  $A_{\mathrm{C}}^{\mathrm{ATLAS7}} = 0.006 \pm 0.010\stat\pm 0.005\syst= 0.006 \pm 0.011\,(\mathrm{stat \oplus syst})$. For the 8\TeV combination, the expected uncertainties after marginalisation are used: the systematic uncertainty from each source is evaluated by building pseudo-data with variations of the predictions and for each one repeating the unfolding procedure, yielding $A_{\mathrm{C}}^{\mathrm{ATLAS8}} = 0.0090 \pm 0.0044\stat\pm 0.0025\syst$. The sum in quadrature of these individually estimated uncertainties is equal to the total uncertainty quoted in Table~\ref{tab:Summary7and8}.

The charge asymmetry is also measured as a function of several kinematic $\ttbar$ variables, including the mass of the $\ttbar$ system.

\subsection{The CMS measurements}

The CMS results used for the combinations are based on data taken at 7\TeV~\cite{CMS_7TeV} and 8\TeV~\cite{CMS_8TeV_Template,CMS_8TeV_Unfolding}, corresponding to integrated luminosities of 5.0 and $19.6\fbinv$, respectively. Two different approaches are pursued for these measurements. The two approaches feature similar criteria to select events with a lepton+jets signature, but differ in the procedures to reconstruct the $\ttbar$ pair and to measure the charge asymmetry. In this paper, the analyses are named as follows: ``CMS unfolding''~\cite{CMS_7TeV,CMS_8TeV_Unfolding} refers to the analysis in which the reconstructed distributions are unfolded to parton level and the charge asymmetry is measured in the unfolded distributions, similar to the approach in the ATLAS analysis, while ``CMS template''~\cite{CMS_8TeV_Template} refers to a measurement of the charge asymmetry by fitting the reconstructed distribution using dedicated templates for symmetric and antisymmetric components of the sensitive observable. The CMS unfolding analysis provides results at both centre-of-mass energies, while the CMS template analysis has only been performed on the data collected at 8\TeV.

For the simulation of the $\ttbar$ signal at the two centre-of-mass energies, the NLO event generator \textsc{powheg-Box} 1.0 is used, together with the CTEQ6M~\cite{Pumplin:2002vw} (for 7\TeV) and CT10 (for 8\TeV) NLO PDF sets. Single-top-quark background events are simulated using the LO event generator \textsc{MadGraph}~\cite{madgraph} 5.1.3 (for 7\TeV) or \textsc{powheg-Box} (for 8\TeV). All events are generated assuming a top quark mass of $m_{\PQt} = 172.5\GeV$. The contributions from $\PW$ and $\Z$ boson production in association with jets are simulated using \textsc{MadGraph}. The parton shower is simulated using \textsc{pythia}~(6.4.24 for 7\TeV and 6.4.26 for 8\TeV)  with the Z2~\cite{Chatrchyan:2011id} tune for 7\TeV and the Z2$^*$~\cite{Khachatryan:2015pea, Chatrchyan:2013gfi} tune for 8\TeV in all samples. Data events from a sideband region, defined by relaxing the isolation or identification criteria for the lepton, are used to model the multijet background in the signal region.

To select events with a lepton+jets signature, candidate events are required to feature one highly energetic electron or muon, well isolated from other activity in the detector, and at least four jets. One of these jets has to be classified as a $\PQb$-jet.
For that purpose, in the 7\TeV analyses, the TCHE algorithm~\cite{csvtagger} is used which is
based on the track with the second-highest impact parameter significance.
At 8\TeV the more sophisticated combined secondary vertex (CSV) tagger~\cite{csvtagger} is applied.
The tagging efficiencies of these algorithms at the chosen working points are 60\% (TCHE) and 70\% (CSV), with a rejection factor for light-quark and gluon jets of about 100. The selection criteria of the two analyses differ in the required minimum $\pt$ of the selected jets. The CMS unfolding analysis considers jets with $\pt>30\GeV$, while in the CMS template analysis all jets with $\pt>20\GeV$ are taken into account. As a result, the number of events available for the CMS template analysis is almost twice the number of events analysed in the CMS unfolding measurement.

\subsubsection{The CMS unfolding analysis}

In this analysis, the charge asymmetry is measured both inclusively and differentially as a function of different variables characterising the $\ttbar$ system. In order to calculate the values of $\Delta\abs{y}$ and the variables used in the differential measurements, the four-momenta of top quarks and antiquarks are reconstructed from the decay products observed in the detector. The leptonically decaying $\PW$ boson is reconstructed from the electron or muon and $\ptvecmiss$. The assignment of the reconstructed jets to the final-state quarks is based on the $\PQb$-tagging information of the jets and the invariant masses of the reconstructed top quarks and the hadronically decaying $\PW$ boson.

To suppress the background contributions from multijet and  $\Z$+jets production, the events are required to satisfy $\pt^{\text{miss}} > 40\GeV$ in the 7\TeV analysis and  $m_{\mathrm{T}}^{\PW} > 50\GeV$ in the 8\TeV analysis. The invariant mass of the three-jet combination that yields the largest transverse momentum, $M_{3}$,
is highly correlated with the mass of the hadronically decaying top quark. The $M_{3}$  variable can therefore be used to distinguish between processes including a top quark and processes that do not contain any top quark. In the  7\TeV  analysis, \ptmiss and $M_{3}$ are fitted simultaneously to estimate the background contributions, while at 8\TeV, the  $m_{\mathrm{T}}^{\PW}$ variable and $M_{3}$ are used. At both centre-of-mass energies, a signal fraction of 80\% is estimated after applying the full selection, with the largest background contribution coming from $\PW$+jets production.

The $\Delta\abs{y}$ distribution, as well as the distributions of the other relevant variables, is corrected for background contributions and migration effects due to the resolution of the reconstruction procedure, and extrapolated from the selected events to the full phase space of $\ttbar$ production. This is done by unfolding with the Tikhonov-based regularised matrix inversion method as implemented in TUnfold~\cite{TUnfold}. The measured inclusive charge asymmetries at the two centre-of-mass energies are listed in Table~\ref{tab:Summary7and8}.

\subsubsection{The CMS template analysis}

In this analysis, the reconstruction of the four-vectors of the top quark and antiquark is performed for all jet-parton assignments. For each assignment, the four-momenta of the selected jets are corrected according to the partons to which they are assigned. Flavour-dependent scale factors correcting the jet energies from reconstruction level to parton level are derived from simulated $\ttbar$ events. In each event, one of the possible assignments is chosen based on a likelihood criterion that uses the $\PQb$-tagging information and the invariant masses of the reconstructed hadronically decaying $\PW$ boson and top quark. The energy resolution of the selected jets corresponding to the chosen assignment is further improved by applying a kinematic fit under the $\ttbar$ hypothesis.

For the determination of the composition of the selected sample, a likelihood
discriminant is constructed from the $m_{\mathrm{T}}^{\PW}$ variable and from the probability that at least one of the possible jet-parton assignments is the correct one. The latter is expressed by the product of two independent probabilities: one based on the invariant masses of the reconstructed hadronically decaying top quark and $\PW$ boson, and one based on the $\PQb$-tagging information of the jets assigned to $\PQb$-quarks and light quarks. A maximum-likelihood fit using the distribution of the resulting discriminant is employed to determine the contributions from $\ttbar$ signal events, $\PW$+jets events, and events from multijet production. The contributions from $\Z$+jets processes and single-top-quark production are fixed to their SM predictions. The signal fraction in the selected data, determined with this method, is about 65\%.

The charge asymmetry is measured in a second maximum-likelihood fit. The sensitive variable used in this analysis is $\Upsilon_{\ttbar} = \tanh{\Delta\abs{y}}$.
The hyperbolic tangent preserves the same asymmetry properties as the canonical $\Delta\abs{y}$, with the additional feature that it is bounded to the range between $-1$ and 1. Signal events, simulated with the \textsc{powheg-Box} event generator, are used to construct the symmetric and antisymmetric components of the probability distribution $\rho(\Upsilon_{\ttbar})$ for the variable $\Upsilon_{\ttbar}$. From a linear combination of these two components a generalised model can be constructed with a single parameter $\alpha$ that varies the amplitude of the antisymmetric component. The sample composition is fixed to the results of the first likelihood fit, and the charge asymmetry is determined by finding the value of $\alpha$ for which the $\Upsilon_{\ttbar}$ model best fits the data.

This method is only used to measure the inclusive charge asymmetry with 8\TeV data, and the result is shown in Table~\ref{tab:Summary7and8}.

\section{Combination method}
\label{sec:blue}

Three separate combinations of ATLAS and CMS results are performed: a combination of the inclusive results at 7\TeV, a combination of the inclusive results at 8\TeV, and a combination of the charge asymmetry measured differentially as a function of the invariant mass of the $\ttbar$ system at 8\TeV. For the combination of the inclusive results at 8\TeV, the two most precise measurements are considered: the ATLAS analysis and the CMS template analysis. Because of the large correlation between the two CMS analyses available at 8\TeV,
the impact of also adding the result of the 8\TeV CMS unfolding analysis to the combination is negligible
(within the quoted precision of $10^{-4}$). For the combination of the differential measurements, the ATLAS result is combined with the result of the CMS unfolding analysis.

In each combination, the two input results are combined by finding the best linear unbiased estimate (BLUE)~\cite{blue1,blue2} with the method implemented in Ref.~\cite{blue_implementation}.
The BLUE method finds the coefficients to be used in a linear combination of the input measurements by minimising the total uncertainty of the combined result, taking into account both the statistical and systematic uncertainties, as well as correlations between the inputs.

\section{Classification of uncertainties and correlation of measurements}
\label{sec:syst}

Dedicated studies are performed to estimate the influence of various systematic uncertainties on the results for all analyses considered for the combinations. In the ATLAS analyses, the expected impact of systematic uncertainties is studied with alternative pseudo-data distributions built from the expected signal and background contributions. These alternative pseudo-data sets are generated by varying each source of systematic uncertainty by one standard deviation (${\pm}1 \sigma$). The unfolding procedure is then repeated using the baseline background templates and response matrices.
At 7\TeV, the impact of the uncertainty on the measured
$A_{\mathrm{C}}$ value is reported as the difference between the mean value of the nominal posterior and the mean value of the varied posterior, and the maximum between the positive and negative variation is chosen as symmetric uncertainty.
At 8\TeV, the average asymmetry variation $|A_{\mathrm{C}}(+1\sigma)-A_{\mathrm{C}}(-1\sigma)|/2$ is quoted as the systematic uncertainty for each source.

In the CMS analyses, the uncertainties are estimated by repeating the analyses
on data using simulated samples modified by the variations under study. This
includes signal and background templates for the estimation of the signal and
background contributions to the selected data set, as well as varied input information for the unfolding and template fitting procedure, respectively. The uncertainty is then defined as the difference between the results obtained using the default simulation and the results obtained using the altered simulation. Where two shifts for the same source of uncertainty are available the absolute value of the larger shift is chosen as the systematic uncertainty.

Although the list of evaluated sources of systematic uncertainty and the
actual procedures to estimate their impacts are partially different between
the analyses, it is still possible to identify contributions that describe
similar physical effects. The classification of the systematic uncertainties
evaluated by the ATLAS and CMS analyses is described in the following, and
their assumed correlations across experiments are given in parentheses. In the description below,
"---" means that the uncertainty source is considered only either in the ATLAS
or the CMS measurement. For some sources of uncertainty, two values are given
for the correlation assumptions. In these cases, the first value applies
to the combination of inclusive results (ATLAS analysis and CMS template
analysis), while the second value applies to the combination of differential results (ATLAS analysis and CMS unfolding analysis). The classification of uncertainties is coarser for the combination of results at 7\TeV. For these analyses, several sources of uncertainty with similar physics origin are grouped together and only the resulting categories are used for the combination. The higher precision of the results at 8\TeV necessitates finer splitting of the various contributions to the overall systematic uncertainty.
For a limited number of correlated systematic sources, it was verified that a particular systematic variation induces a change in each measurement in the same direction. The correlation of the measurements for each correlated uncertainty source is assumed to be always positive. The stability of the combinations under variations on the assumptions about the correlations between the systematic uncertainties is checked in Section~\ref{sec:stability}. The potential impact on the combination from using different procedures to estimate the systematic uncertainties in the ATLAS and CMS analyses is studied, for selected uncertainty contributions, by varying the correlations and the sizes of these systematic uncertainties. The combination is found to be stable against such variations. Uncertainty sources whose effect on the combined central value is smaller than the quoted precision, i.e. smaller than 0.001 for the combination at 7\TeV and smaller than 0.0001 for the combination at 8\TeV, are considered negligible.

\subsection{Uncertainties at \texorpdfstring{7\TeV}{7 TeV}}
\label{sec:syst7TeV}

\begin{itemize}
\item{\textbf{Statistical uncertainty in data}} (Correlation: 0) \\
This category includes the statistical uncertainty in the resulting asymmetry
due to the size of the available data set.
\item {\textbf{Statistical uncertainty in simulation}} (Correlation: 0)\\
In each analysis, the impact of the statistical uncertainty in the extraction of the response matrix from simulation, used for unfolding, is estimated by independently varying the
individual elements of the response matrix within their statistical uncertainties.
\item {\textbf{Detector model}} (Correlation: 0)\\
  For each simulated event, scale factors are applied depending on the kinematic properties of the leptons to correct for their mismodelling in the simulation. The uncertainties in these scale factors are propagated to the asymmetry measurement. In addition, the uncertainty from lepton charge misidentification is included in the CMS analysis. Scale factors are also applied to the simulated jets to correct for small differences between the efficiencies for $\PQb$-jet identification in simulation and data. The impact of the corresponding uncertainties on the measurements is found to be negligible in both analyses. All sources of uncertainty coming from the jet energy calibration, as well as from the modelling of the jet energy resolution in the simulation, are also considered as detector-related uncertainties. All these sources of systematic uncertainties that are related to the modelling of the detectors are grouped together and assumed to be uncorrelated.  Some smaller components of these uncertainties in the jet energy scale (JES) are, however, known to be correlated between the two experiments~\cite{JESATLASCMS7TeV}. As their impact is negligible compared to the uncorrelated components, the assumption of no correlation between the two experiments is still appropriate.

\item {\textbf{Pile-up and missing transverse momentum}} (Correlation: 0)\\
  In both analyses, additional interactions in the same or nearby bunch crossings (pile-up events) are overlaid on the simulated signal and background events. The distribution of pile-up events is adjusted by taking into account the measured instantaneous luminosities per bunch and an effective $\Pp\Pp$ inelastic cross section. The impact on the CMS analysis from the uncertainty in this adjustment is found to be negligible. In the ATLAS analysis, this uncertainty is taken into account in the calculation of \ptmiss, in addition to propagated uncertainties in lepton and jet momenta.

\item {\textbf{Signal modelling}} (Correlation: 0.5)\\
In both analyses, the impact of using different MC event generators (\textsc{powheg-Box} instead of \textsc{alpgen} in the ATLAS analysis, and
\textsc{MadGraph} instead of \textsc{powheg-Box} in the CMS analysis) is
studied. Also, the difference between the hadronisation and parton shower models
implemented in \textsc{pythia} and \textsc{herwig} interfaced with the event generators described
above is taken as a systematic uncertainty. Variations by a factor 0.5 and 2.0 of the factorisation and renormalisation
scales in matrix element calculations, as well as in the parton shower, are also included. The assumed correlations for the different uncertainty sources of this category range from anticorrelated to fully correlated, thus a correlation of 50\% is applied in the combination. However, since the impact of these uncertainties on the ATLAS measurement is negligible and taken to be zero in the combination, the actual choice of the correlation value does not affect the result.

\item {\textbf{PDF}} (Correlation: 1)\\
The systematic uncertainties arising from the choice of PDFs used in the simulation are estimated in both analyses using the eigenvectors of the CTEQ6.6~\cite{CTEQ6.6} PDF set. In the ATLAS analysis, two additional sets of PDFs are taken into account (MSTW2008~\cite{MSTW08} and NNPDF2.1~\cite{NNPDF2.1}) in the estimation of the PDF uncertainty spread, as described in Ref.~\cite{PDF4LHC}.
In both analyses, the uncertainty is evaluated for the signal events, as well as for the main background source, $\PW$+jets events.

\item{\textbf{Backgrounds}}
\begin{itemize}
\item {\textit{Modelling of multijet production}} (Correlation: 0)\\
In the ATLAS analysis, both the shape and normalisation uncertainties in the multijet background
are estimated through the matrix method~\cite{MM}. The impact of this uncertainty on the asymmetry measurement
is found to be negligible. In the CMS analysis, effects of shape variations on
the $\Delta\abs{y}$ template from the multijet background model based on data are examined by inverting for each event the sign of the
$\Delta\abs{y}$ value to get an alternative shape.

\item {\textit{Modelling of $\PW$+jets}} (Correlation: 0.5)\\
In the ATLAS analysis, the \textsc{alpgen} factorisation and renormalisation scales are varied
in the $\PW$+jets simulation with the same factors as for the signal model uncertainty.
In the CMS analysis, a different approach is pursued: the
templates for $\PW$+jets events with negatively charged leptons are interchanged
with the templates for $\PW$+jets events with positively charged
leptons. As the two methods are only partially correlated, a 50\% correlation is assumed between these two systematic uncertainties. As shown in Section~\ref{sec:stability}, this particular choice of correlation strength has no significant impact on the combination result.
\end{itemize}

\item {\textbf{Model dependence}} (Correlation: ---)\\
In both analyses, the potential dependence on the underlying model used to develop the analysis is estimated by using signal samples corresponding to different physical models. The two approaches differ, however, in the type and range of the evaluated variations in the asymmetry. In the ATLAS analysis, relatively small variations in the asymmetry are tested, resulting from different physics models beyond the standard model used for the simulation of the signal sample (\textit{specific physics models}). In the CMS analysis, on the other hand, relatively large variations are tested, regardless of whether or not their realisation is possible in any physics model (\textit{general simplified models}). Because of these differences in the range and motivation of the tested variations, the resulting uncertainties are treated as two different uncertainties, each only evaluated in one of the analyses.
\end{itemize}

\subsection{Uncertainties at \texorpdfstring{8\TeV}{8 TeV}}
\label{sec:syst8TeV}

\begin{itemize}

\item {\textbf{Statistical uncertainty in data}}~(Correlation: 0)\\
For the ATLAS analysis and the CMS template analysis, this category includes the statistical uncertainty in the resulting asymmetries. In the ATLAS measurement, the statistical uncertainty is obtained by running the unfolding procedure without taking any systematic effects into account. For the CMS unfolding analysis, this category includes, in addition, the uncertainties in the normalisation of the background processes and the uncertainty due to the number of simulated events in the simulated signal samples. The uncertainties in the normalisation of the background processes and the uncertainty due to the number of simulated events are treated as systematic uncertainties in the ATLAS analysis and the CMS template analysis. The statistical uncertainties are considered to be uncorrelated between the ATLAS and CMS analyses.

\item{\textbf{Statistical uncertainty in simulation}} ~(Correlation: 0 / ---) \\
 In the ATLAS analysis, the impact of the size of the simulated signal event
 sample is investigated by repeating the unfolding several times on one single
 pseudo-data set with randomly drawn response matrices. A similar approach is
 followed in the CMS unfolding analysis, with the only differences being that
 the unfolding is carried out on the measured data instead of one pseudo-data set, and that the resulting uncertainty is included in the statistical uncertainty of the result. The CMS template analysis uses ensembles of alternative templates, generated by varying the original templates according to Poisson statistics to estimate the impact of the sample size. This systematic uncertainty is considered to be uncorrelated between the ATLAS analysis and the CMS template analysis.

\item{\textbf{Detector model (excluding JES)}}
\begin{itemize}
\item{\textit{Leptons}}~(Correlation: 0) \\
  This category includes the uncertainty in the scale factors that are used to correct for mismodelling of the trigger response, as well as of the lepton identification and reconstruction efficiency. The resulting uncertainties are considered to be uncorrelated between the ATLAS and CMS analyses.

\item{\textit{Jet energy resolution}}~(Correlation: 0) \\
This category includes the contributions due to uncertainties in the modelling of the jet energy resolution. These uncertainties are considered to be uncorrelated between the ATLAS and CMS analyses.

\item{\textit{$\PQb$-tagging}}~(Correlation: 0) \\
The two collaborations split up the different sources of $\PQb$-tagging-related systematic uncertainties in a different way. However, as the $\PQb$-tagging uncertainties in all three analyses are relatively small contributions to the overall uncertainty, the different $\PQb$-tagging uncertainties for each analysis are combined by adding their contributions in quadrature to obtain only one $\PQb$-tagging uncertainty per analysis. The resulting uncertainties are assumed to be uncorrelated between the ATLAS and CMS analyses.

\item{\textit{Missing transverse momentum}}~(Correlation: ---) \\
The \ptmiss reconstruction is affected by uncertainties associated with leptons, jet energy scales and resolutions, which
are propagated to the \ptmiss calculation, as well as by uncertainties associated with the modelling of the underlying
event. This was estimated only in the ATLAS analysis and the effect is small; for the CMS analyses, the uncertainties due to the underlying event modelling are studied and found to be negligible compared to other \ptmiss-related uncertainties. The effects on the \ptmiss value are, however, not estimated as a separate source of uncertainty but considered implicitly in other uncertainty sources by propagating variations into the calculation of \ptmiss. Thus, no assumption about the correlation between ATLAS and CMS has to be made.

\item{\textit{Pile-up}}~(Correlation: ---) \\
In the CMS analyses, the impact of a potential mismodelling of the pile-up component overlaid on the simulated events is estimated by varying the number of simulated pile-up events. For the ATLAS analysis, this uncertainty contribution is included in one of the jet energy scale uncertainties described below (\textit{Uncorrelated JES}), and thus not treated as a separate source of uncertainty. Again, no assumption about the correlation between ATLAS and CMS has to be made.

\end{itemize}

\item {\textbf{Jet energy scale}}\\
In all three analyses, the same jet clustering algorithm, the anti-\kt algorithm~\cite{Cacciari:2008gp}, is employed, with different values for the radius parameter $R$ ($R = 0.4$ for ATLAS, $R = 0.5$ for CMS). In the ATLAS analysis, jets are built from energy deposits in the calorimeter, while in the CMS analyses, jets are reconstructed from particle-flow~\cite{PF} objects. Thus, the calibration procedure and uncertainties in the calibration are quite different. Both collaborations determine a large number of single sources that contribute to the overall uncertainty in the jet energy calibration, which can be grouped into categories with similar ranges of correlation~\cite{JESATLASCMS}.

\begin{itemize}

\item {\textit{Uncorrelated} JES}~(Correlation: 0)  \\
This category includes all components of the JES uncertainties that are uncorrelated between ATLAS and CMS (i.e. statistical and detector-related effects, pile-up, and high-$\pt$ uncertainty components) and uncertainties that do not match between the two experiments.

\item {\textit{Partially correlated} JES}~(Correlation: 0.5) \\
  This category includes the modelling uncertainties of the \textit{in situ} methods used in both collaborations for the determination of the jet energy calibration. While both collaborations measure similar systematic effects, the actual impact has some dependence on the detector itself and on the choices of the technical implementation. The uncertainties of the two analyses are therefore assumed to be partially correlated and the range from 0.0 to 0.5 is scanned. The most conservative result is obtained assuming a correlation of 50\%.

\item {\textit{Mostly correlated} JES}~(Correlation: 1) \\
The calibration of jet energies across $\eta$ with respect to a well-understood central reference region is sensitive to the modelling of radiation. In both collaborations, similar event generators and techniques are used to estimate the influence of different radiation patterns in the simulation on the jet energy. These uncertainties are expected to be highly correlated, although there are some expected decorrelations from differences in the analysis procedures. As such, the uncertainties assigned to this category are treated as mostly correlated and the range from 0.5 to 1.0 is evaluated. The most conservative result is obtained assuming a correlation of 100\%.

\item {\textit{Fully correlated} JES}~(Correlation: 1) \\
Uncertainties in the JES due to the flavour of jets are grouped into this category. This includes effects due to differences in the jet energy response for various jet flavours and effects from differences in the flavour mixture, compared to the mixture used in the calibration procedures. The assumed correlation between ATLAS and CMS for this uncertainty category is 100\%.

\end{itemize}

\item{\textbf{Signal modelling}}
\begin{itemize}
\item{\textit{Generator}} ~(Correlation: 1) \\
The impact of using a different MC event generator is studied in all three analyses: \textsc{mc}$@$\textsc{nlo} instead of \textsc{powheg-Box} in the ATLAS analysis,  \textsc{MadGraph} instead of \textsc{powheg-Box} in the CMS unfolding analysis, and \textsc{mc}$@$\textsc{nlo} instead of \textsc{powheg-Box} in the CMS template analysis. These uncertainties are assumed to be fully correlated between the ATLAS and CMS analyses.

\item{\textit{Parton shower and hadronisation}}~(Correlation: ---/1) \\
The results with different implementations of parton showering and hadronisation in \textsc{pythia} and \textsc{herwig}, interfaced with the event generators listed above, are used to estimate the uncertainties in the modelling of the parton shower and hadronisation process. This uncertainty category is assumed to be 100\% correlated between the ATLAS analysis and the CMS unfolding analysis, while it has not been evaluated for the CMS template analysis. The impact of an additional uncertainty for the CMS template analysis on the combination was studied by using the parton shower and hadronisation uncertainty of the CMS unfolding analysis, and was found to be small (see Section~\ref{sec:stability8tev}).

\item{\textit{Initial-state and final-state radiation and choice of the factorisation and renormalisation scales}}~(Correlation: 1) \\
  The uncertainties due to the modelling of additional radiation are assumed to be fully correlated between the ATLAS and CMS analyses,
although the two collaborations follow different methods to estimate this impact. In the ATLAS analysis, the uncertainty associated with the modelling of initial-state and final-state radiation is estimated using the \textsc{acermc} event generator with varied parameters.  In the CMS analyses, a different approach is followed and the factorisation and renormalisation scales are varied in the simulation of the matrix element, as well as in the parton shower, by factors of 0.5 and 2.0.

\item{\textit{Modelling of the top quark $\pt$}} ~(Correlation: ---) \\
  In the CMS analyses, the simulated $\ttbar$~events are reweighted to correct the $\pt$ spectrum of the generated top quarks, which was found to be significantly harder than the observed spectra in differential cross section measurements~\cite{Khachatryan:2015oqa}. To estimate the uncertainty related to this mismodelling, the measurement is repeated without the reweighting and the difference with respect to the default result is taken as a measure of the uncertainty. In the ATLAS analysis, the measured $\pt$ distribution of the reconstructed top quarks is found to agree well with the simulation using the \textsc{powheg-Box}+\textsc{herwig} MC sample~\cite{ATLAS_diffxsec8TeV}. Since this sample is used to derive the uncertainty due to parton shower and hadronisation modelling, no additional uncertainty is assigned for the modelling of the $\pt$ spectrum of top quarks in the ATLAS analysis. For the CMS template analysis, the contribution of this uncertainty is found to be negligible.

\item{\textit{Parton distribution functions}}~(Correlation: 1) \\
  The uncertainties due to the choice of PDFs are considered to be fully correlated between ATLAS and CMS analyses. In the ATLAS analysis and CMS unfolding analysis, the uncertainty in the PDFs is estimated using three different PDF sets (CT10, MSTW2008, and NNPDF2.1), while in the CMS template analysis the variations of the CT10 eigenvectors are used and variations of the strong coupling parameter are considered independently. In the two CMS analyses, the influence of PDF variations on the modelling of the $\ttbar$ process is estimated, while in the ATLAS analysis the impact of the choice of PDF on the modelling of the $\PW$+jets background is also evaluated. The latter is captured in the two CMS analyses by the uncertainty source ``Modelling of $\PW$+jets'' in the category ``Backgrounds''.

\end{itemize}

\item{\textbf{Integrated luminosity}} ~(Correlation: ---) \\
 The luminosity used to construct the fit model in the CMS template analysis is varied by $\pm$4.4\%. In the other two analyses,
varying the integrated luminosity has negligible effect on the results.

\item{\textbf{Backgrounds}}
\begin{itemize}
\item{\textit{Single-top-quark and $\Z$+jets}}~(Correlation: 1/---) \\
In the ATLAS analysis, an uncertainty arises from the uncertainty in the single-top-quark and $\Z$+jets background normalisations from the theoretical cross sections, as well as an additional normalisation uncertainty for each additional jet. In the CMS template analysis, this uncertainty includes the uncertainties in the single-top-quark and $\Z$+jets production cross sections, as well as the uncertainty in the ratio of single-top-quark to single-top-antiquark production. In the CMS unfolding analysis, the uncertainties in the normalisation of background contributions are considered when subtracting these contributions from the measured data distribution in the unfolding step and are therefore contained in the statistical uncertainty of the result and not treated as a separate systematic uncertainty. For the combination of the ATLAS analysis and CMS template analysis, these uncertainties are considered fully correlated, while they do not apply to the CMS unfolding analysis.

\item{\textit{Modelling of multijet production}}~(Correlation: 0) \\
In the ATLAS analysis, the uncertainties in the multijet background come from the size of the data sample used in the matrix method, as well
as from the uncertainties in the estimation of the misidentified lepton probability in different control regions. In the CMS template analysis, the uncertainty comes mainly from the size of the data sideband sample used for the modelling of the multijet process. In the CMS unfolding analysis, variations in the shapes of the multijet templates derived from sideband regions in data, are used to estimate the uncertainty in the modelling of this background component. These uncertainties are considered to be uncorrelated between the ATLAS and CMS analyses.

\item{\textit{Modelling of $\PW$+jets}}~(Correlation: ---) \\
In order to estimate the influence of a possible mismodelling of the simulated
$\PW$+jets background, the CMS unfolding measurement is repeated using a
$\PW$+jets template determined from a sideband region in data, defined by
requiring exactly zero $\PQb$-tagged jets. The CMS template analysis constructs
an alternative $\PW$+jets template by increasing or decreasing the number of
events with heavy-flavour jets. In addition, the uncertainty in the estimated
$\PW$+jets normalisation is propagated into this uncertainty. The $\PW$+jets
component of the data set is estimated \textit{in situ} in the ATLAS analysis using separate templates for the different flavours, and therefore no additional uncertainty is taken into account.
\end{itemize}

\item{\textbf{Method}} (Correlation: ---/0) \\
  The uncertainty from the unfolding procedure is determined using pseudo-data
  sets in which the $\ttbar$ signal events are reweighted to simulate different values of the charge asymmetry by applying the default response matrix and comparing the true charge asymmetry values with the values obtained after unfolding. In the ATLAS analysis, a linear fit of the generator-level asymmetry versus the unfolded asymmetry is performed and the uncertainty in the unfolding is calculated from the slope and the offset of this fit.
  In the CMS unfolding analysis, different scenarios for reweighted
  $\ttbar$~events are examined and the differences between the unfolded
  asymmetries and the reweighted true asymmetries are taken to be a measure of
  the model dependence of the unfolding procedure. The uncertainty is assumed
  to be uncorrelated between ATLAS measurement and the CMS unfolding
  measurement. In the CMS template analysis, pseudo-data sets are constructed for different values of $A_{\mathrm{C}}$ by changing the mixture of symmetric and antisymmetric components in the sample of simulated signal events. The difference between the mean charge asymmetry measured for each alternative model and the actual charge asymmetry is found to be negligible.

\end{itemize}

Even if the systematic uncertainties for the 7 and 8\TeV results are not fully evaluated in the same way, most of the assigned correlations between the ATLAS and CMS measurements are identical between 7 and 8\TeV.
As discussed in Section~\ref{sec:syst7TeV}, the systematic uncertainty in the detector model is assumed to be uncorrelated at 7\TeV since the components that could be considered correlated (for example JES) are small. For the signal modelling uncertainties, the correlation choice of 0.5 at 7\TeV does not affect the result because the signal modelling has a negligible effect on the ATLAS measurement at 7\TeV.

\subsection{Correlation assumptions in the combination of differential results at \texorpdfstring{8\TeV}{8 TeV}}
\label{sec:DiffCorrelations}

An additional complication in the combination of differential measurements, with respect to inclusive ones, is that the correlations between different bins in the mass distribution of the $\ttbar$ system have to be taken into account in addition to the correlations between categories of systematic uncertainties. The ATLAS analysis and the CMS unfolding analysis use the same number of bins and the same ranges for the bins. The mapping of systematic uncertainties in each bin between the ATLAS analysis and the CMS unfolding analysis is the same as for the combination of the inclusive results.

Bin-to-bin correlations are considered by the analyses for each
uncertainty. In the CMS unfolding analysis, the statistical correlations are calculated \textit{in situ} during the unfolding process. To evaluate the bin-to-bin correlations for systematic uncertainties, for each source of systematic uncertainty u, the measurement is repeated on data using modified simulated samples, resulting in new values for the charge asymmetries per bin. The changes in unfolded asymmetries are then used to construct a systematic covariance matrix in a loose analogy to statistical covariance matrices.
For an uncertainty described by a single systematic shift, a covariance of \\
\begin{linenomath}
\begin{equation}
 \mathrm{cov}^{\mathrm{u}}(c_{i},c_{j}) = \left (A_{\mathrm{C}}^{\mathrm{u}}(c_{i})-A_{\mathrm{C}}^{\mathrm{nom}}(c_i) \right ) \, \left ( A_{\mathrm{C}}^{\mathrm{u}}(c_j) - A_{\mathrm{C}}^{\mathrm{nom}}(c_j) \right )
\label{eq:cmscov}
\end{equation}
\end{linenomath}
is used, with $c_{i}$ and $c_{j}$ referring to bins in $m_{\ttbar}$ of the CMS analysis, and $A_{\mathrm{C}}^{\mathrm{u}}(c_i)$ and $A_{\mathrm{C}}^{\mathrm{u}}(c_j)$ being the asymmetry values for these bins resulting from the systematic shift, and $A_{\mathrm{C}}^{\mathrm{nom}}(c_i)$ and $A_{\mathrm{C}}^{\mathrm{nom}}(c_j)$ being the results of the nominal measurement.

For uncertainties that are determined using exactly two variations (up and down variation, indexed by $+$ and $-$), the absolute values of the maximal shifts observed in each result bin are determined by
\begin{linenomath}
\begin{equation*}
\Delta_{\mathrm{max}} A_{\mathrm{C}}(c_i) = \mathrm{max}\left (\left|A_{\mathrm{C}}^{\mathrm{u},+}(c_i) - A_{\mathrm{C}}^{\mathrm{nom}}(c_i)\right |, \left |A_{\mathrm{C}}^{\mathrm{u},-}(c_i) - A_{\mathrm{C}}^{\mathrm{nom}}(c_i)\right | \right),
\end{equation*}
\end{linenomath}
and the covariance is then defined as
\begin{linenomath}
\begin{equation*}
 \mathrm{cov}^{\mathrm{u}}(c_i,c_j) = \Delta_{\mathrm{max}} A_{\mathrm{C}}(c_i) \, \Delta_{\mathrm{max}} A_{\mathrm{C}}(c_j) \, \mathrm{sign}\left [\left ( A_{\mathrm{C}}^{\mathrm{u},+}(c_i) - A_{\mathrm{C}}^{\mathrm{u},-}(c_i)\right ) \, \left (A_{\mathrm{C}}^{\mathrm{u},+}(c_j)-A_{\mathrm{C}}^{\mathrm{u},-}(c_j) \right )\right ].
\end{equation*}
\end{linenomath}
This procedure corresponds to a symmetrisation of the largest observed shifts, and thus constitutes a more conservative uncertainty estimate than an approach based on a direct analogy with statistical covariance definitions. The resulting covariance matrices are transformed into correlation matrices for the combination. For uncertainty sources that are described by a single systematic shift, the covariance matrix in Eq.~(\ref{eq:cmscov}) corresponds to a correlation matrix with ${\pm}1$ entries only:
\begin{linenomath}
\begin{equation}
 \corr^{\mathrm{u}}(c_{i},c_{j}) = \mathrm{sign}\left [\left (A_{\mathrm{C}}^{\mathrm{u}}(c_{i})-A_{\mathrm{C}}^{\mathrm{nom}}(c_i) \right ) \, \left ( A_{\mathrm{C}}^{\mathrm{u}}(c_j) - A_{\mathrm{C}}^{\mathrm{nom}}(c_j) \right ) \right ] .
\label{eq:cmscorr}
\end{equation}
\end{linenomath}
However, for uncertainties where several shifts are combined, e.g. the
uncertainty from the PDF choice, first all covariance matrices are combined, and this combined covariance matrix is then transformed into a correlation matrix, yielding entries in the correlation matrix different from ${\pm}1$.

In the ATLAS analysis, the correlations between bins $a_i$ and $a_j$ for each source of systematic uncertainty
that is marginalised are extracted by projecting the posterior probability density into the $(a_i,a_j)$ plane.
The correlation is then computed from the ensemble points that build the probability density in this plane, taking the average between the up and down variations as for the computation of the systematic uncertainty itself. The correlation matrix for the modelling uncertainties that are not marginalised
is obtained using Eq.~(\ref{eq:cmscorr}), yielding entries with ${\pm}1$. The correlation matrix for the
systematic uncertainty coming from the size of the simulated samples is taken to be the identity matrix.

For the estimation of the correlations between bins from two different experiments that use the same binning scheme, one can distinguish two cases: the correlations between the same bins of the ATLAS and CMS results, and the correlations between different bins of the ATLAS and CMS results. In the first case, for each uncertainty source, the correlation assumptions of the inclusive combination are used.

The bin-to-bin correlations for a given uncertainty can be different in the two analyses because of the differences in the unfolding techniques and in the methods used to determine these correlations. This is especially the case for uncertainty categories that include several contributions, e.g. the uncertainty from the PDF choice or the JES uncertainty categories. A straightforward determination of the correlation between different bins in the ATLAS and CMS measurements is therefore not possible. These off-diagonal elements of the correlation matrices are instead calculated by multiplying the assumed correlation of the uncertainty u, $\rho_{\mathrm{u}}$, with a correction factor depending on the bin-to-bin correlations within the individual experiments.

The correlation for a given uncertainty source u between $a_{i}$ and $c_{j}$ can thus be calculated as:
\begin{linenomath}
\begin{equation}
\corr^{\mathrm{u}}(a_{i}, c_{j})' = \corr^{\mathrm{u}}(a_{i}, c_{i})\,\corr^{\mathrm{u}}(c_{i}, c_{j}) = \rho_{\mathrm{u}}\,\corr^{\mathrm{u}}(c_{i}, c_{j}).
\label{eq:corrCC}
\end{equation}
Due to symmetry reasons, there is also an alternative way to calculate the correlation, starting from the ATLAS bin-to-bin-correlations instead of the CMS ones:
\begin{equation}
\label{eq:corrAA}
\corr^{\mathrm{u}}(a_{i}, c_{j})'' = \corr^{\mathrm{u}}(a_{i}, a_{j})\,\corr^{\mathrm{u}}(a_{j}, c_{j}) = \corr^{\mathrm{u}}(a_{i}, a_{j})\cdot\rho_{\mathrm{u}}.
\end{equation}
As both methods are valid, the mean of the two estimates $\corr^{\mathrm{u}}(a_{i}, c_{j})'$ and $\corr^{\mathrm{u}}(a_{i}, c_{j})''$ is used for the combination
\begin{equation}
\corr^{\mathrm{u}}(a_{i}, c_{j}) = \rho_{\mathrm{u}}\,\frac{\corr^{\mathrm{u}}(c_{i}, c_{j}) + \corr^{\mathrm{u}}(a_{i}, a_{j})}{2},
\label{eq:corraverage}
\end{equation}
\end{linenomath}
while the impact of using either only Eq.~(\ref{eq:corrCC}) or (\ref{eq:corrAA}) is studied as a check of the stability of the combination.

\begin{table}[b!]
\topcaption{Uncertainties in the input measurements, assumed correlations $\rho$ between the uncertainties, and the resulting values for the uncertainties in the inclusive combination at $\sqrt{s} = 7\TeV$.
Systematic uncertainties smaller than 0.001 are shown as ``${<}0.001$" in the table and are ignored in the combination.}
\centering
\begin{tabular}{lrrrr}
\hline
                                & ATLAS    & CMS      & $\rho$ & Combined  \\
\hline
$A_{\mathrm{C}}$                & 0.006    & 0.004    & 0.058   & 0.005\\
\hline
 Statistical (data)                    & 0.010    & 0.010    & 0  & 0.007 \\
 \hline
 Statistical (simulation)                 & 0.002    & 0.002    & 0  & 0.001 \\
 Detector model                 & 0.004    & 0.007    & 0  & 0.004 \\
 Pile-up+\ptmiss   & 0.002    & $<$0.001 & 0  & 0.001 \\
 Signal modelling               & $<$0.001 & 0.002    & 0.5  & 0.001 \\
 PDF                            & 0.001    & 0.002    & 1  & 0.001 \\
 Multijet                       & $<$0.001 & 0.001    & 0  & $<$0.001 \\
 $\PW$+jets                       & 0.002    & 0.004    & 0.5& 0.003 \\
 Model dependence               &          &          &    &   \\
   ~~~Specific physics models   & $<$0.001 &  ---     & ---  & $<$0.001 \\
   ~~~General simplified models &    ---   & 0.007    & ---  & 0.002 \\
\hline
Systematic uncertainty          & 0.005    & 0.011    &    & 0.006\\
Total uncertainty               & 0.011    & 0.015    &    & 0.009\\
\hline
\end{tabular}

\label{tab:Syst7TeV}
\end{table}

\section{Combination of the results}
\label{sec:results}

\subsection{Combination of inclusive results at \texorpdfstring{7\TeV}{7 TeV}}

\begin{figure}[ht]
\centering
\includegraphics[width=0.8\textwidth]{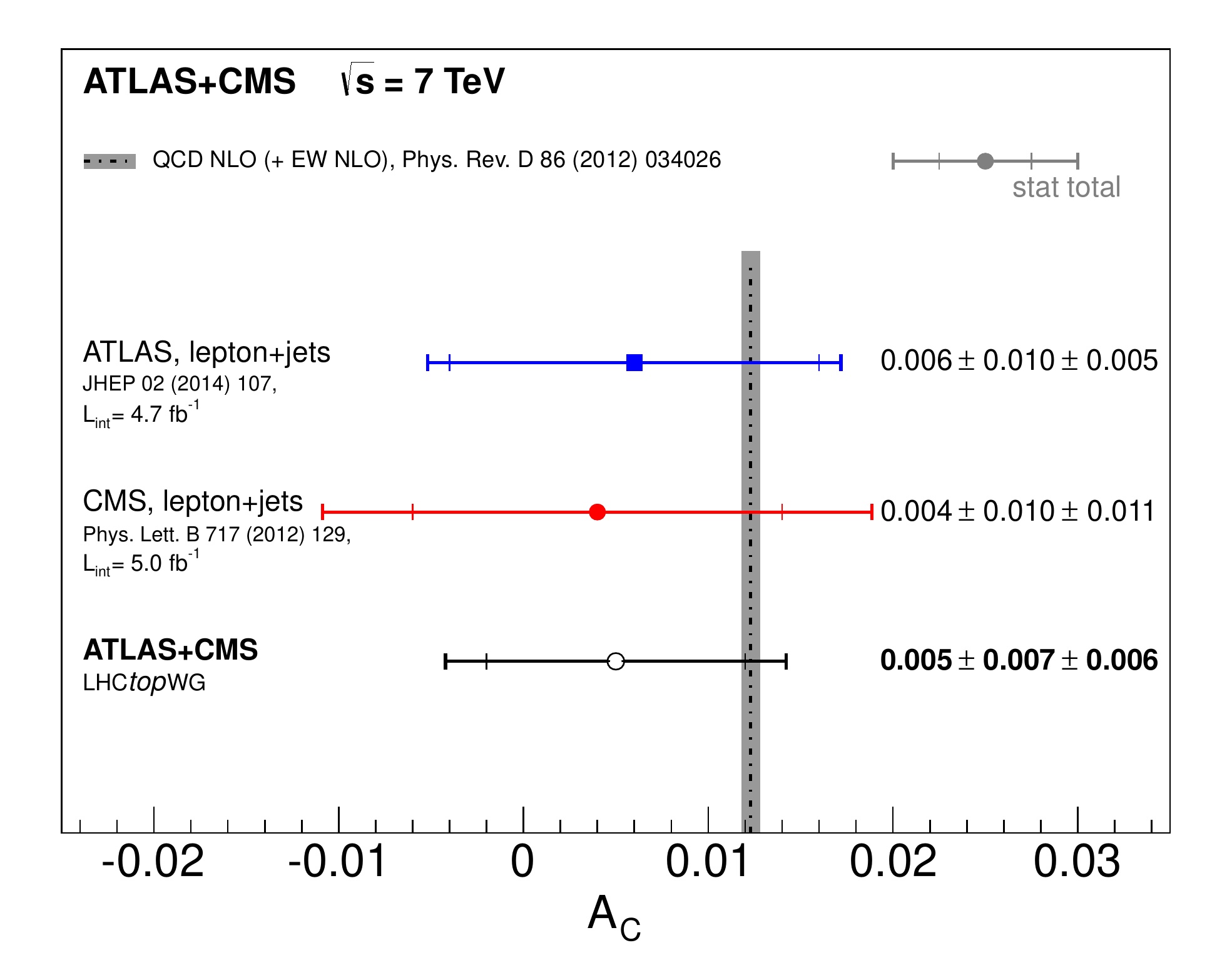}

\caption{Summary of the single inclusive measurements and the LHC combination
  at $\sqrt{s} = 7\TeV$ compared to the theoretical prediction
  calculated at NLO precision in the strong coupling constant (including NLO electroweak corrections). The inner bars indicate the statistical uncertainty, while the outer bars indicate the total uncertainty. The uncertainty in the theoretical prediction is dominated by uncertainties due to scale variations.}
\label{fig:inclusivecombination7}
\end{figure}

The value resulting from the combination of the inclusive measurements at 7\TeV is
\begin{linenomath}
\begin{equation*}
A_{\mathrm{C}}^{\mathrm{LHC7}} = 0.005 \pm 0.007\stat \pm 0.006\syst.
\end{equation*}
\end{linenomath}
This corresponds to an improvement in the total uncertainty of about 18\% with respect to the ATLAS result alone, and an improvement of about 40\% with respect to the CMS result.
The ATLAS result contributes with a weight of 0.65 to the combined result, while the weight of the CMS result is 0.35.
The $\chi^{2}$ of the combination is 0.012  (one degree of freedom), corresponding to a $p$-value of 0.91. A breakdown of the individual uncertainties and their combined values is provided in Table~\ref{tab:Syst7TeV}. Figure~\ref{fig:inclusivecombination7} illustrates the two individual inclusive results and their combination, along with the respective uncertainties, and compares them to the prediction from theory~\cite{Bernreuther}, calculated at NLO, including electroweak (EW) corrections.

\subsection{Combination of inclusive results at \texorpdfstring{8\TeV}{8 TeV}}

\begin{figure}[ht]
\centering
\includegraphics[width=0.8\textwidth]{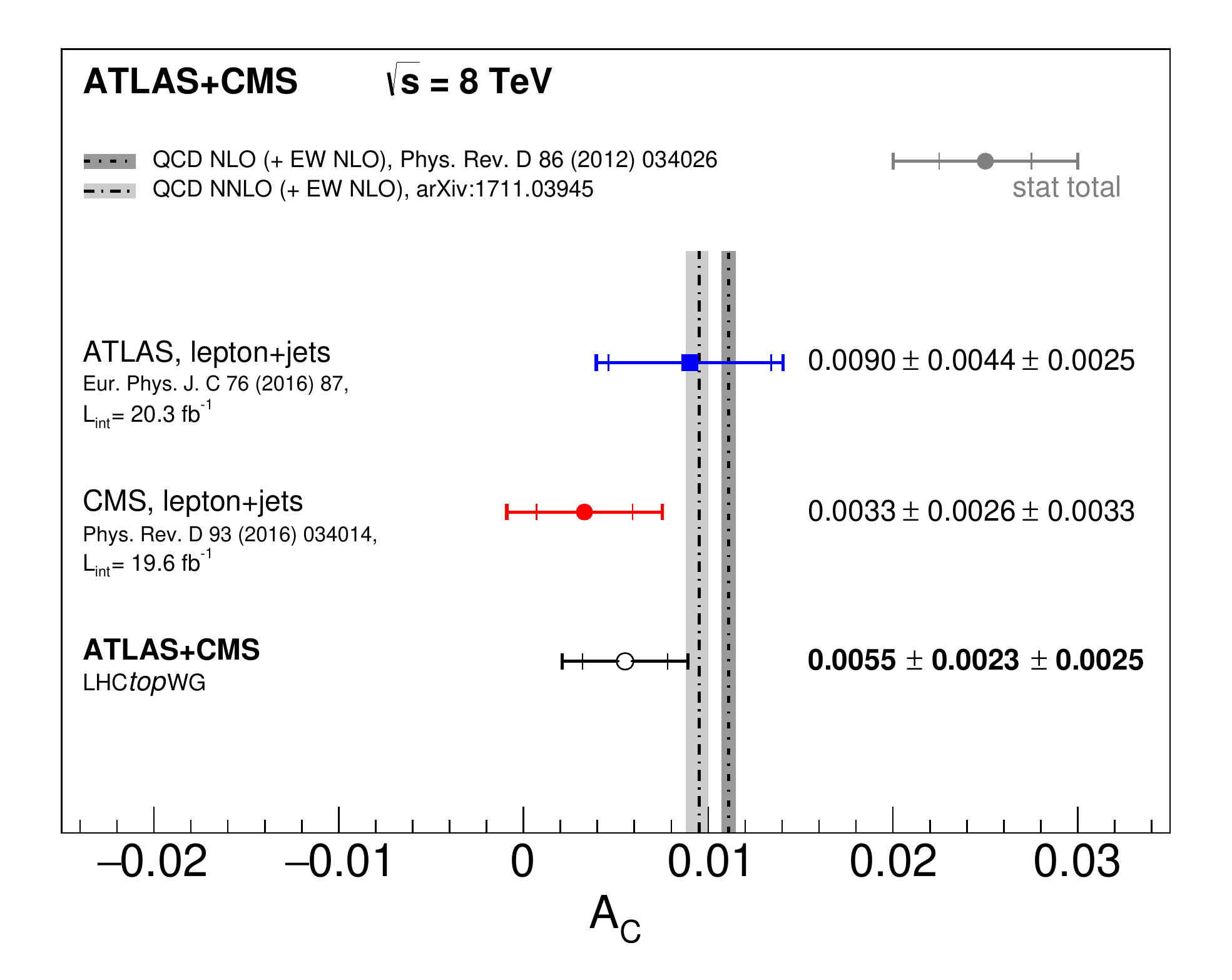}

\caption{Summary of the single inclusive measurements and the LHC combination at $\sqrt{s} = 8\TeV$ compared to theoretical predictions at NLO~\cite{Bernreuther} and NNLO~\cite{nnlo1} precision in the strong coupling constant (including NLO electroweak corrections). The inner bars indicate the statistical uncertainty, while the outer bars indicate the total uncertainty. The uncertainty in the theoretical predictions is dominated by uncertainties due to scale variations.}
\label{fig:inclusivecombination8}
\end{figure}

\begin{figure}[ht]
\centering
\includegraphics[width=0.6\textwidth]{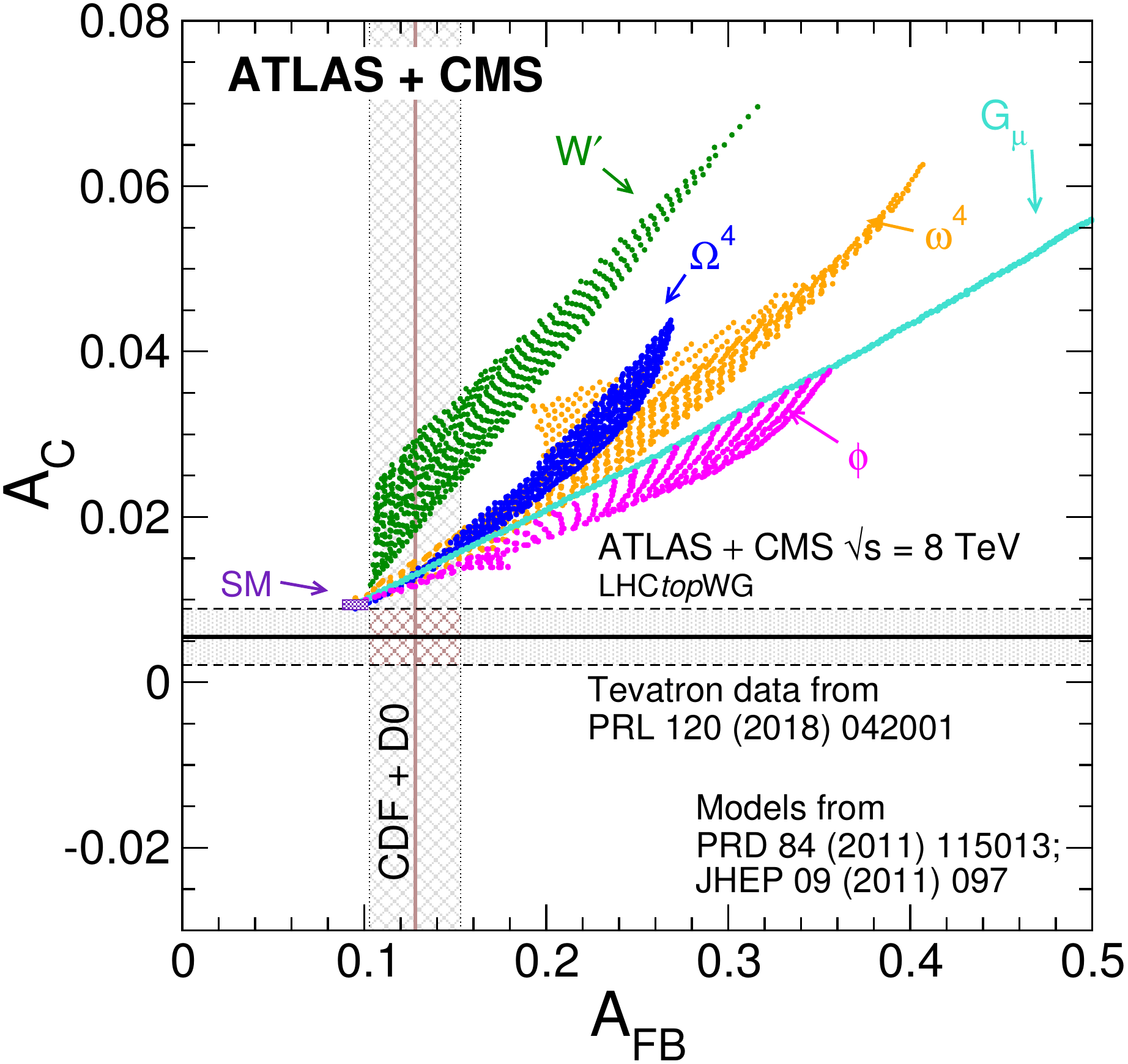}

\caption{Measured inclusive charge asymmetry $A_{\mathrm{C}}$ at the LHC at $\sqrt{s} = 8\TeV$ (horizontal line) versus forward-backward asymmetry $A_{\mathrm{FB}}$ (vertical lines)
at the Tevatron~\cite{Aaltonen:2017efp}, compared with the SM prediction at QCD NNLO (+EW NLO)~\cite{nnlo2,nnlo3,nnlo4} and predictions incorporating various potential BSM contributions~\cite{AguilarSaavedra:2011hz, AguilarSaavedra:2011ug}: a $\PWpr$ boson, a heavy axigluon ($G_{\mu}$), a scalar isodoublet ($\phi$), a colour triplet scalar ($\omega^{4}$), and a colour sextet scalar ($\Omega^{4}$).}
\label{fig:modelkiller}
\end{figure}

The resulting value of the combination of the inclusive measurements at 8\TeV is
\begin{linenomath}
\begin{equation*}
A_{\mathrm{C}}^{\mathrm{LHC8}} = 0.0055 \pm 0.0023\stat \pm 0.0025\syst.
\end{equation*}
\end{linenomath}
This corresponds to an improvement in the total uncertainty of about 32\% with respect to the ATLAS result alone, and an improvement of about 17\% with respect to the CMS template result.
The ATLAS result contributes with a weight of 0.39 to the combined result, while the weight of the CMS template
result is 0.61.
The $\chi^{2}$ of the combination is 0.88  (one degree of freedom), corresponding to a $p$-value of 0.35.
 A breakdown of the individual uncertainties and their combined values is provided in Table~\ref{tab:Syst8TeV}. Figure~\ref{fig:inclusivecombination8} illustrates the two individual inclusive results and their combination, along with the respective uncertainties, and compares them to theoretical predictions.
  Figure~\ref{fig:modelkiller} compares the $A_{\mathrm{C}}$ value from the combination described in this article and the  $A_{\mathrm{FB}}$ values measured by the Tevatron experiments with predictions from the SM and from BSM theories (see~\cite{Aguilar-Saavedra:2014kpa} for a review). The BSM theories include models with charged $\PWpr$ bosons with right-handed couplings, heavy colour-octet vector gluons $G_{\mu}$ with axial couplings, colour-singlet Higgs boson like isodoublets $\phi$, colour-triplet scalars $\omega^{4}$, and colour-sextet scalars $\Omega^{4}$ with right-handed flavour-violating $tu$ couplings. Details of these BSM models can be found in Refs.~\cite{AguilarSaavedra:2011hz,AguilarSaavedra:2011ug}.
The combined $A_{\mathrm{C}}$ value uniquely restricts wide regions of the possible BSM parameter space, e.g. for axigluon models.

\begin{table}[h!]
\topcaption{Uncertainties in the input measurements, assumed correlations $\rho$ between the uncertainties, and the resulting values for the uncertainties in the inclusive combination at $\sqrt{s} = 8\TeV$. The breakdown of the systematic uncertainties of the ATLAS analysis (for the systematic uncertainties that are marginalised) corresponds to the expected uncertainties after marginalisation.}
\centering
\begin{tabular}{lrrrr}
\hline
& ATLAS & CMS  &  $\rho$ & Combined   \\
\hline
  $A_{\mathrm{C}}$                     &  0.0090     &   0.0033  &    0.13        & 0.0055\\
\hline

Statistical (data)                   &  0.0044     &   0.0026  &     0      & 0.0023 \\
\hline
Statistical (simulation)        & 0.0010  & 0.0015 & 0 &  0.0010 \\
\hline
\textit{Detector model (excluding JES)} & & & & \\
Leptons			      & 0.0003 & 0.0001 & 0 & 0.0001 \\
Jet energy resolution         & 0.0005 & 0.0004 & 0 & 0.0003 \\
$\PQb$-tagging                     & 0.0004 & 0.0007 & 0 & 0.0005\\
Missing transverse momentum   & 0.0002 & --- & ---  & 0.0001\\
Pile-up                        & ---     & 0.0003 & --- & 0.0002\\
\hline
\textit{Jet energy scale}  & & & & \\
Uncorrelated JES              & 0.0010 & 0.0004 & 0 & 0.0005\\
Partially correlated JES      & 0.0009 & 0.0010 & 0.5 & 0.0008\\
Mostly correlated JES         & 0.0002 & 0.0004 & 1 & 0.0003\\
Fully correlated JES          & 0.0009 & 0.0008 & 1 & 0.0008\\
\hline
\textit{Signal modelling} & & & & \\
Event generator                    & 0.0004 & 0.0002 & 1 & 0.0003\\
Parton shower and hadronisation  & 0.0004 & --- & --- & 0.0002\\
Scale/radiation              & 0.0009     & 0.0014 & 1 & 0.0012\\
PDF                          & 0.0007  & 0.0002 & 1 & 0.0004\\
\hline
Integrated luminosity         & --- & 0.0001 & --- & 0.0001\\
\hline
\textit{Backgrounds}  & & & & \\
Single-top-quark / $\Z$+jets         & 0.0001 & 0.0004 & 1 & 0.0003\\
Multijet                 & 0.0005 & 0.0018 & 0 & 0.0011\\
$\PW$+jets                       & --- & 0.0002 & --- & 0.0001\\
\hline
Method                         & 0.0003 & --- & --- & 0.0001\\
\hline
Systematic uncertainty       &  0.0025 & 0.0033  &  & 0.0025\\
Total uncertainty            &  0.0051 & 0.0041  &  & 0.0034\\
\hline
\end{tabular}

\label{tab:Syst8TeV}
\end{table}

\subsection{Combination of differential results at \texorpdfstring{8\TeV}{8 TeV}}

\begin{figure}[t]
\centering
\includegraphics[width=0.7\textwidth]{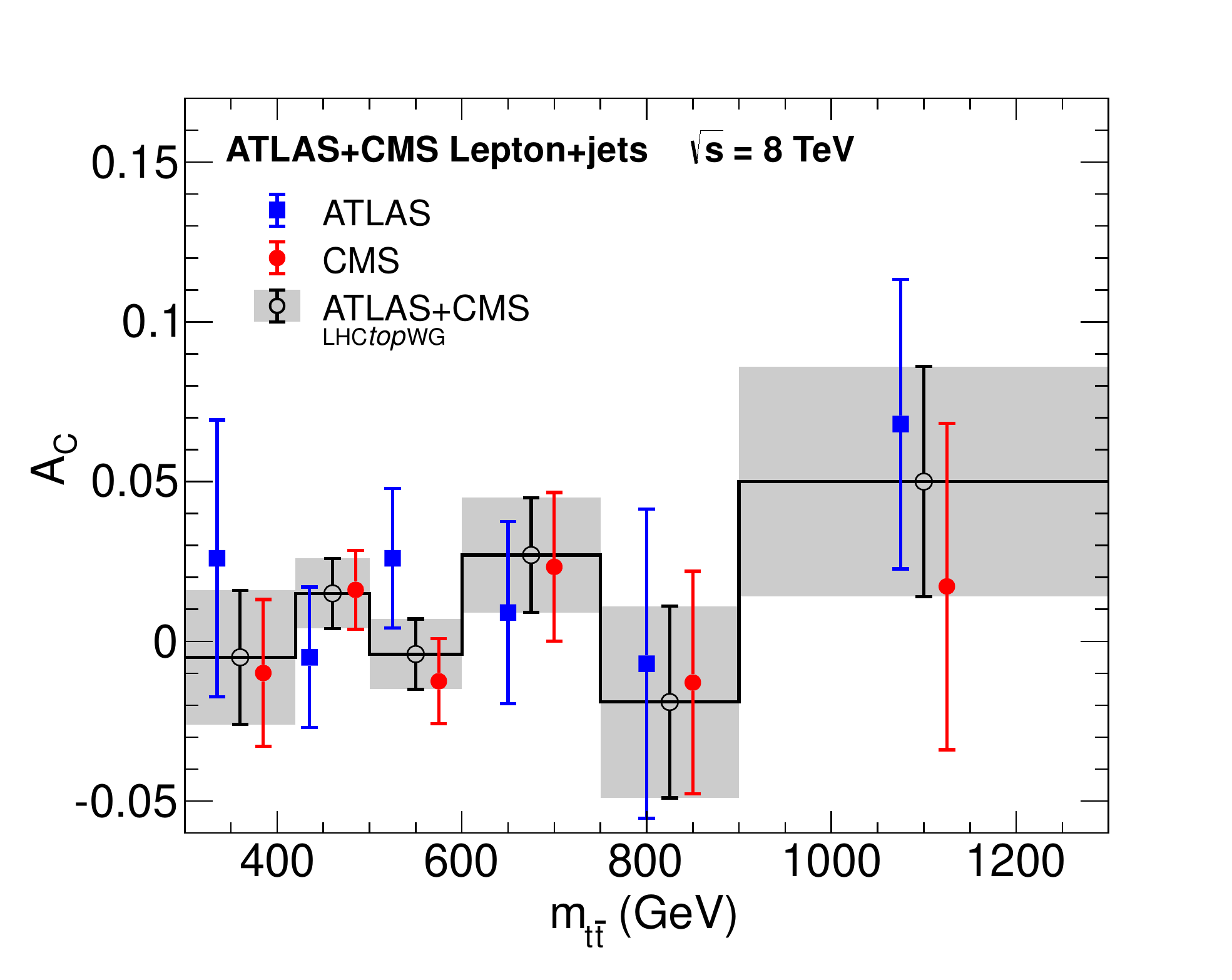} \\

\caption{Charge asymmetry in six bins of the invariant mass of the $\ttbar$ system as measured in the ATLAS and CMS analyses and the combined results. The last bin includes the overflow. The gray band indicates the uncertainty in the combined result.}
\label{fig:diffcombination}
\end{figure}

\begin{figure}[t]
\centering
\includegraphics[width=0.7\textwidth]{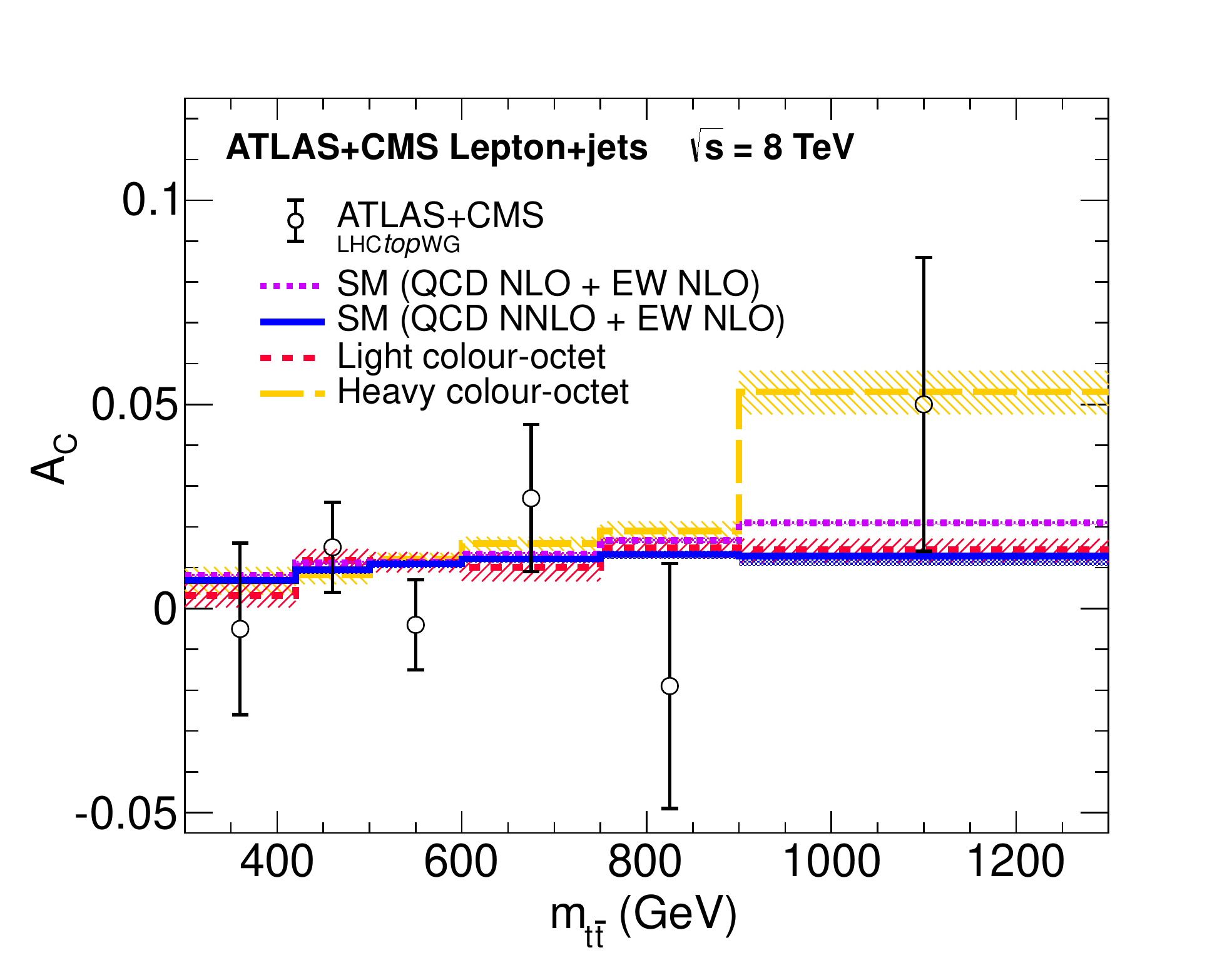}

\caption{The combined ATLAS+CMS charge asymmetry in six bins of the invariant mass of the $\ttbar$ system in comparison with theoretical predictions for the SM~\cite{Bernreuther,nnlo1} and two versions of a colour-octet model~\cite{Aguilar-Saavedra:2014nja}. The last bin includes the overflow, both for the combination and the theory predictions. The uncertainties, indicated by the shaded areas, reported for the SM predictions are dominated by scale variations and are small, while the uncertainties reported for the colour-octet model are statistical uncertainties in the simulation.}
\label{fig:diffcombination2}
\end{figure}

The values of the combined charge asymmetries in six bins of $m_{\ttbar}$ can be found in Table~\ref{tab:BlueResDiff}. Within their statistical and systematic uncertainties, the combined results agree with the predictions from SM calculations.
Depending on the $m_{\ttbar}$ bin, the total uncertainty corresponds to an improvement in precision between 20\% (bin 6) and 52\% (bin 1) with respect to the ATLAS measurement alone, and to an improvement in precision between 9\% (bin 1) and 31\% (bin 6) with respect to the CMS measurement alone. Table~\ref{tab:CorRes} shows the correlations between the bins of the combined result. Looking only at the leading contributions, the ATLAS measurement contributes, depending on the $m_{\ttbar}$ bin, with a weight between 0.22 (bin 1) and 0.59 (bin 6) to the combined results, while the weights of the CMS results lie in the range from 0.41 (bin 6) to 0.78 (bin 1). The total $\chi^{2}$ of the combination is 4.01  (six degrees of freedom), corresponding to a $p$-value of 0.69.
Figure~\ref{fig:diffcombination} shows the differential $A_{\mathrm{C}}$ distributions of the individual measurements and the combination, along with their total uncertainties.
In Figure~\ref{fig:diffcombination2}, the combined result is compared to two predictions for the SM, calculated at NLO~\cite{Bernreuther} and at NNLO~\cite{nnlo2, nnlo3,nnlo4}, and to predictions for two versions of a colour-octet model~\cite{Aguilar-Saavedra:2014nja}.
The latter models are examples that give moderate positive contributions to the forward-backward asymmetry at the Tevatron and a negligible contribution to the inclusive charge asymmetry at the LHC. However, they differ significantly in the asymmetry for the last bin with the difference being similar in size to the uncertainty in the combined LHC result.

\begin{table}[h!t]
\topcaption{ATLAS and CMS charge asymmetry results at 8\TeV in six bins of $m_{\ttbar}$ and the combined values along with statistical and systematic uncertainties. In addition the predictions from QCD calculations at NLO~\cite{Bernreuther}
and NNLO~\cite{nnlo1} are given.}
\centering
\newcolumntype{.}{D{.}{.}{2.5}}
\begin{tabular}{l......}
\hline
&\multicolumn{6}{c}{$m_{\ttbar}$ bin (\GeVns{})}\\
                        &  \multicolumn{1}{c}{$<$420} & \multicolumn{1}{c}{420--500} & \multicolumn{1}{c}{500--600} & \multicolumn{1}{c}{600--750} & \multicolumn{1}{c}{750--900} & \multicolumn{1}{c}{$>$900} \\
\hline
ATLAS $A_{\mathrm{C}}$  &  0.026   & -0.005    & 0.026     & 0.009  &  -0.007    & 0.068  \\
statistical uncertainty              &  0.025 & 0.017 & 0.018 & 0.023 & 0.042 & 0.037 \\
systematic uncertainty              &  0.036 & 0.015 & 0.012  & 0.017 & 0.025 & 0.026 \\
\hline
CMS $A_{\mathrm{C}}$    &  -0.010  &  0.016    & -0.013    & 0.023     & -0.013    & 0.017 \\
statistical uncertainty              & 0.020 & 0.011 & 0.012 & 0.015 & 0.026 & 0.038 \\
systematic uncertainty              & 0.012 & 0.006 & 0.006 & 0.018 & 0.023 & 0.035\\
\hline
Combined $A_{\mathrm{C}}$ &  -0.005     & 0.015       & -0.004      & 0.027       & -0.019      & 0.050 \\
statistical uncertainty              & 0.017  & 0.009  & 0.009  & 0.012  & 0.022  & 0.027 \\
systematic uncertainty              & 0.013  & 0.006  & 0.007  & 0.014  & 0.020  & 0.024 \\
\hline
QCD NLO + EW NLO          & 0.00809     &  0.01117    &   0.01138   &  0.01335    &   0.01671   &  0.02100     \\
uncertainty             & 0.00036 & 0.00053 & 0.00039 & 0.00038 & 0.00058 & 0.00025 \\
QCD NNLO + EW NLO        & 0.00690     & 0.00950     &   0.01095   &  0.01219    &  0.01327    & 0.01286  \\
uncertainty             & +0.00061    & +0.00081    & +0.00079    & +0.00054   & +0.00075    & +0.00083 \\
                        & -0.00058    & -0.00087   & -0.00095     & -0.00081   & -0.00101  & -0.00231 \\
\hline
\end{tabular}

\label{tab:BlueResDiff}
\end{table}

\clearpage

\begin{table}[ht]
\topcaption{Correlation matrix for the  six correlated bins of the combined result.}
\centering
\begin{tabular}{lrrrrrr}\hline
Bins & 1 & 2  & 3  & 4  & 5 & 6 \\ \hline
1 & 1.000  &        &       &       &       &       \\
2 & $-$0.129 & 1.000  &       &       &       &       \\
3 & $-$0.100 & 0.015  & 1.000 &       &       &       \\
4 & $-$0.018 & $-$0.060 & $-$0.098& 1.000 &       &       \\
5 & $-$0.009 & $-$0.074 & 0.067 & $-$0.342& 1.000 &       \\
6 & $-$0.088 & $-$0.040 & 0.012 & 0.199 & $-$0.188& 1.000 \\ \hline
\end{tabular}

\label{tab:CorRes}
\end{table}

\section{Stability studies}
\label{sec:stability}

The stability of the combinations with respect to the assumptions made about the correlations between the systematic uncertainties is studied by varying the input assumptions.
The resulting central value and total uncertainty are then compared to the result obtained using the default settings.

\subsection{Stability checks at \texorpdfstring{7\TeV}{7 TeV}}
\label{sec:stability7tev}

At $\sqrt{s} = 7\TeV$, for uncertainties with an assumed default correlation of 100\% or of zero, the assumption of 50\% correlation is tried. For the uncertainty in the modelling of the W$+$jets background, for which the nominal assumption is 50\% correlation, both zero and 100\% correlation are tested. Changing the correlation assumptions in the described way has neither an effect on the central value of the combination nor on its total uncertainty within the quoted precision ($10^{-3}$).

\subsection{Stability checks at \texorpdfstring{8\TeV}{8 TeV}}
\label{sec:stability8tev}

Several studies are performed to assess the stability of the inclusive and differential combinations at $\sqrt{s} = 8\TeV$. In a first check, the combinations are repeated taking only the statistical uncertainties into account to investigate the impact of systematic uncertainties on the combination results. The variations in the central values of the inclusive and differential combinations, and thus the impact of systematic uncertainties, are below $0.5\sigma_{\text{stat}}$, where $\sigma_{\text{stat}}$ is the statistical uncertainty of the central values. Therefore, the impact of the assumptions about the correlations between the sources of systematic uncertainties can be expected to also be modest in size.

As in the combination at 7\TeV, the correlation assumptions are varied one by one and the results are compared with the nominal results. These studies are performed for the combination of the inclusive measurements as well as for the combination of the differential measurements. For the inclusive combination, none of the described changes in the correlation assumption has an impact on the central value of the combination or the total uncertainty larger than the quoted precision ($10^{-4}$). The differential combination is also found to be robust against such changes. The impacts, if any, on the central values or the total uncertainties of single bins are found to be below 0.1$\sigma_{\text{tot}}$, where $\sigma_{\text{tot}}$ is the total uncertainty in that bin.

\subsubsection{Additional stability checks for the differential combination}

Two additional studies are performed for the combination of the differential results.
The first study concerns the different ways of building the correlation matrix in the ATLAS analysis and evaluates their impact on the central values and uncertainties of the combination results.
In the ATLAS analysis, the bin-to-bin correlations for all systematic uncertainties that are not marginalised are set to $\pm$1, and the correlations for systematic uncertainties that are marginalised are derived directly from the unfolding.
This nominal method to construct the correlation matrix is compared to a scenario where the correlations are all set to $+1$, and a scenario where the correlations for the marginalised uncertainties are set to either $+1$ or $-1$ depending on whether the value of the correlation derived from the unfolding takes positive or negative values. These checks of ignoring the correlations extracted from the unfolding represent a conservative test for the case where it
would not be known how to compute these correlations.
The differences in the central values between the nominal scenario and these two different approaches are below $0.4\sigma_{\text{tot}}$ in each $m_{\ttbar}$ bin, where $\sigma_{\text{tot}}$ is the total uncertainty in that bin. The impact on the total uncertainties in all six bins is at most 0.002.

In the second study, the procedure to estimate the bin-to-bin inter-experiment correlations as defined in Eq.~(\ref{eq:corraverage}) is changed to using the bin-to-bin correlations either only from ATLAS or only from CMS:
\begin{equation*}
\corr^{u}(a_{i}, c_{j}) = \rho_{u}\, \corr^{u}(c_{i}, c_{j}),\quad\text{or}
\end{equation*}
\begin{equation*}
\corr^{u}(a_{i}, c_{j}) = \rho_{u}\, \corr^{u}(a_{i}, a_{j}).
\end{equation*}

Using only the bin-to-bin correlations from the ATLAS (CMS) analysis changes the central values by at most $0.3\sigma_{\text{tot}}$ ($0.2\sigma_{\text{tot}}$), with $\sigma_{\text{tot}}$ being the total uncertainty for the respective bin. The total uncertainties remain stable or increase by no more than 0.002 in the case of using only the bin-to-bin correlations from the ATLAS analysis. Overall, the changes in the central values in all the bins are well inside the range given by the total uncertainty in each bin.

\subsubsection{Impact of the parton shower and hadronisation systematic uncertainty in the inclusive combination}

The CMS template analysis does not take into account an uncertainty due to the modelling of parton showers and hadronisation. As a further test, the impact of such an additional uncertainty on the inclusive combination results is estimated by using the corresponding uncertainty values from the CMS unfolding analysis (0.0011), scaled by a factor of 2 as an approximation. The factor 2 is motivated by looking at the ratios of the other signal-modelling-related uncertainties between the two CMS analyses. In this study, the central value of the combination increases by 0.16$\sigma_{\text{tot}}$, and the total uncertainty increases by 12\% (relative). These variations are small and similar in size to variations obtained in the other stability tests.

\section{Summary}
Combinations of ATLAS and CMS measurements of the $\ttbar$ charge asymmetry in lepton+jets final states are reported using proton-proton collision data collected at the LHC at centre-of-mass energies of 7 and 8\TeV and corresponding to integrated luminosities for each experiment of up to 5 and $20\fbinv$, respectively. Inclusive charge asymmetry results are combined at 7 and 8\TeV, and differential measurements of the charge asymmetry as a function of the invariant mass of the $\ttbar$ system at 8\TeV are combined.
Detailed studies of the correlations between the different measurements and systematic uncertainties have been performed.
The precision of the resulting combinations is significantly improved with respect to the corresponding individual measurements. The individual results and the combinations are in agreement with standard model calculations at next-to-leading-order and next-to-next-to-leading-order precision and also compatible with zero asymmetry. They uniquely restrict the phase space of possible new physics phenomena which would produce asymmetries larger than the standard model ones.

\begin{acknowledgments}
We thank CERN for the very successful operation of the LHC, as well as the support staff from our institutions without whom ATLAS and CMS could not be operated efficiently. The work presented in this paper is performed in the context of the LHC top physics working group and benefits from its important contributions.

We acknowledge the support of ANPCyT (Argentina); YerPhI (Armenia); ARC (Australia); BMWFW and FWF (Austria); ANAS (Azerbaijan); SSTC (Belarus); FNRS and FWO (Belgium); CNPq, CAPES, FAPERJ, and FAPESP (Brazil); MES (Bulgaria); NSERC, NRC, and CFI (Canada); CERN; CONICYT (Chile); CAS, MoST, and NSFC (China); COLCIENCIAS (Colombia); MSES and CSF (Croatia); RPF (Cyprus); MSMT CR, MPO CR, and VSC CR (Czech Republic); DNRF and DNSRC (Denmark); SENESCYT (Ecuador); MoER, ERC IUT, and ERDF (Estonia); Academy of Finland, MEC, and HIP (Finland); CEA and CNRS/IN2P3 (France); SRNSF (Georgia); BMBF , DFG, HGF, and MPG (Germany); GSRT (Greece); RGC (Hong Kong SAR, China); OTKA and NIH (Hungary); DAE and DST (India); IPM (Iran); SFI (Ireland); ISF, I-CORE, and Benoziyo Center (Israel); INFN (Italy); MEXT and JSPS (Japan); MSIP, and NRF (Republic of Korea); LAS (Lithuania); MOE and UM (Malaysia); BUAP, CINVESTAV, CONACYT, LNS, SEP, and UASLP-FAI (Mexico); CNRST (Morocco); NWO (Netherlands); MBIE (New Zealand); RCN (Norway); PAEC (Pakistan); MNiSW, MSHE, NCN, and NSC (Poland); FCT (Portugal); MNE/IFA (Romania); JINR (Dubna); MES, MON, NRC KI, RosAtom, RAS, RFBR and RAEP (Russian Federation); MESTD (Serbia); MSSR (Slovakia); ARRS and MIZ\v{S} (Slovenia); DST/NRF (South Africa); MINECO, SEIDI, CPAN, PCTI and FEDER (Spain); SRC and Wallenberg Foundation (Sweden); ETH Board, ETH Zurich, PSI, SERI, SNSF, UniZH, and Cantons of Bern, Geneva and Zurich (Switzerland); MOST (Taipei); ThEPCenter, IPST, STAR, and NSTDA (Thailand); TUBITAK and TAEK (Turkey); NASU and SFFR (Ukraine); STFC (United Kingdom); DOE and NSF (United States of America).

In addition, individual groups and members have received support from BELSPO, FRIA, and IWT (Belgium); BCKDF, the Canada Council, CANARIE, CRC, Compute Canada, FQRNT, and the Ontario Innovation Trust (Canada); the Leventis Foundation (Cyprus); MEYS (Czech Republic); EPLANET, ERC, ERDF, FP7, Horizon 2020, and Marie Sk{\l}odowska-Curie Actions (European Union); Investissements d'Avenir Labex and Idex, ANR, R{\'e}gion Auvergne and Fondation Partager le Savoir (France); AvH Foundation (Germany); the Herakleitos, Thales, and Aristeia programmes co-financed by EU-ESF and the Greek NSRF (Greece); CSIR (India); BSF, GIF, and Minerva (Israel); BRF (Norway); the HOMING PLUS programme of the FPS, co-financed from the EU Regional Development Fund, the Mobility Plus programme of the MSHE, and the OPUS programme of the NSC (Poland); the NPRP by Qatar NRF (Qatar); Generalitat de Catalunya, Generalitat Valenciana, and the Programa Clar\'in-COFUND del Principado de Asturias (Spain); the Rachadapisek Sompot Fund for Postdoctoral Fellowship, Chulalongkorn University, and the Chulalongkorn Academic into Its 2nd Century Project Advancement Project (Thailand); the Royal Society and Leverhulme Trust (United Kingdom); the A.\ P.\ Sloan Foundation, the Welch Foundation, and the Weston Havens Foundation (United States of America).

The crucial computing support from all WLCG partners is acknowledged gratefully, in particular from CERN and the Tier-1 facilities at TRIUMF (Canada), NDGF (Denmark, Norway, Sweden), CC-IN2P3 (France), KIT/GridKA (Germany), INFN-CNAF (Italy), NL-T1 (Netherlands), RRC-KI and JINR (Russian Federation), PIC (Spain), ASGC (Taipei), RAL (UK), and BNL and FNAL (USA), and from the Tier-2 facilities worldwide and large non-WLCG resource providers.
\end{acknowledgments}

\bibliography{auto_generated}

\providecommand{\href}[2]{#2}\begingroup\raggedright\begin{thebibliography}{10}%
\makeatletter
\providecommand{\hrefCMSnoop }[0]{\@secondoftwo}%
\makeatother
\providecommand{\doi}{\texttt{doi:}\begingroup \urlstyle{tt}\Url}

\bibitem{CDF}
\hrefCMSnoop {}{{CDF} Collaboration, ``{Measurement of the top quark
  forward-backward production asymmetry and its dependence on event kinematic
  properties}'',} \textit{ Phys. Rev. D} \textbf{ 87} (2013) 092002,
  \href{http://dx.doi.org/10.1103/PhysRevD.87.092002}{\doi{10.1103/PhysRevD.87.092002}},
\href{http://www.arXiv.org/abs/1211.1003}{\texttt{arXiv:1211.1003}}.

\bibitem{Dzero}
\hrefCMSnoop {}{{D0} Collaboration, ``{Measurement of the forward-backward
  asymmetry in top quark-antiquark production in $p\bar{p}$ collisions using
  the lepton+jets channel}'',} \textit{ Phys. Rev. D} \textbf{ 90} (2014)
  072011,
  \href{http://dx.doi.org/10.1103/PhysRevD.90.072011}{\doi{10.1103/PhysRevD.90.072011}},
\href{http://www.arXiv.org/abs/1405.0421}{\texttt{arXiv:1405.0421}}.

\bibitem{Aaltonen:2017efp}
\hrefCMSnoop {}{{CDF and D0 Collaborations}, ``{Combined Forward-Backward
  Asymmetry Measurements in Top-Antitop Quark Production at the Tevatron}'',}
  \textit{ Phys. Rev. Lett.} \textbf{ 120} (2018) 042001,
  \href{http://dx.doi.org/10.1103/PhysRevLett.120.042001}{\doi{10.1103/PhysRevLett.120.042001}},
\href{http://www.arXiv.org/abs/1709.04894}{\texttt{arXiv:1709.04894}}.

\bibitem{Kuhn}
\hrefCMSnoop {}{J.~H. K{\"u}hn and G.~Rodrigo, ``{Charge asymmetries of top
  quarks at hadron colliders revisited}'',} \textit{ JHEP} \textbf{ 01} (2012)
  063,
  \href{http://dx.doi.org/10.1007/JHEP01(2012)063}{\doi{10.1007/JHEP01(2012)063}},
\href{http://www.arXiv.org/abs/1109.6830}{\texttt{arXiv:1109.6830}}.

\bibitem{Aguilar-Saavedra:2014kpa}
\hrefCMSnoop {}{J.~A. Aguilar-Saavedra, D.~Amidei, A.~Juste, and
  M.~Perez-Victoria, ``{Asymmetries in top quark pair production at hadron
  colliders}'',} \textit{ Rev. Mod. Phys.} \textbf{ 87} (2015) 421,
  \href{http://dx.doi.org/10.1103/RevModPhys.87.421}{\doi{10.1103/RevModPhys.87.421}},
\href{http://www.arXiv.org/abs/1406.1798}{\texttt{arXiv:1406.1798}}.

\bibitem{Kuhn:1998jr}
\hrefCMSnoop {}{J.~H. K{\"u}hn and G.~Rodrigo, ``{Charge asymmetry in
  hadroproduction of heavy quarks}'',} \textit{ Phys. Rev. Lett.} \textbf{ 81}
  (1998) 49,
  \href{http://dx.doi.org/10.1103/PhysRevLett.81.49}{\doi{10.1103/PhysRevLett.81.49}},
\href{http://www.arXiv.org/abs/hep-ph/9802268}{\texttt{arXiv:hep-ph/9802268}}.

\bibitem{Aad:2008zzm}
\hrefCMSnoop {}{{ATLAS Collaboration}, ``{The ATLAS Experiment at the CERN
  Large Hadron Collider}'',} \textit{ JINST} \textbf{ 3} (2008) S08003,
\href{http://dx.doi.org/10.1088/1748-0221/3/08/S08003}{\doi{10.1088/1748-0221/3/08/S08003}}.

\bibitem{Chatrchyan:2008aa}
\hrefCMSnoop {}{{CMS Collaboration}, ``{The CMS experiment at the CERN LHC}'',}
  \textit{ JINST} \textbf{ 3} (2008) S08004,
\href{http://dx.doi.org/10.1088/1748-0221/3/08/S08004}{\doi{10.1088/1748-0221/3/08/S08004}}.

\bibitem{ATLAS_7TeV}
\hrefCMSnoop {}{{ATLAS Collaboration}, ``{Measurement of the top quark pair
  production charge asymmetry in proton-proton collisions at $\sqrt{s}$ = 7 TeV
  using the ATLAS detector}'',} \textit{ JHEP} \textbf{ 02} (2014) 107,
  \href{http://dx.doi.org/10.1007/JHEP02(2014)107}{\doi{10.1007/JHEP02(2014)107}},
\href{http://www.arXiv.org/abs/1311.6724}{\texttt{arXiv:1311.6724}}.

\bibitem{CMS_7TeV}
\hrefCMSnoop {}{{CMS Collaboration}, ``{Inclusive and differential measurements
  of the $t \bar{t}$ charge asymmetry in proton-proton collisions at $\sqrt{s}
  =$ 7 TeV}'',} \textit{ Phys. Lett. B} \textbf{ 717} (2012) 129,
  \href{http://dx.doi.org/10.1016/j.physletb.2012.09.028}{\doi{10.1016/j.physletb.2012.09.028}},
\href{http://www.arXiv.org/abs/1207.0065}{\texttt{arXiv:1207.0065}}.

\bibitem{ATLAS_8TeV}
\hrefCMSnoop {}{{ATLAS Collaboration}, ``{Measurement of the charge asymmetry
  in top-quark pair production in the lepton-plus-jets final state in $pp$
  collision data at $\sqrt{s}=8\,\mathrm{TeV}$ with the ATLAS detector}'',}
  \textit{ Eur. Phys. J. C} \textbf{ 76} (2016) 87,
  \href{http://dx.doi.org/10.1140/epjc/s10052-016-3910-6}{\doi{10.1140/epjc/s10052-016-3910-6}},
\href{http://www.arXiv.org/abs/1509.02358}{\texttt{arXiv:1509.02358}}.

\bibitem{CMS_8TeV_Template}
\hrefCMSnoop {}{{CMS Collaboration}, ``{Measurement of the charge asymmetry in
  top quark pair production in $pp$ collisions at $\sqrt{s} =$ 8 TeV using a
  template method}'',} \textit{ Phys. Rev. D} \textbf{ 93} (2016) 034014,
  \href{http://dx.doi.org/10.1103/PhysRevD.93.034014}{\doi{10.1103/PhysRevD.93.034014}},
\href{http://www.arXiv.org/abs/1508.03862}{\texttt{arXiv:1508.03862}}.

\bibitem{CMS_8TeV_Unfolding}
\hrefCMSnoop {}{{CMS Collaboration}, ``{Inclusive and differential measurements
  of the $t\overline{t}$ charge asymmetry in $pp$ collisions at $\sqrt{s} =$ 8
  TeV}'',} \textit{ Phys. Lett. B} \textbf{ 757} (2016) 154,
  \href{http://dx.doi.org/10.1016/j.physletb.2016.03.060}{\doi{10.1016/j.physletb.2016.03.060}},
\href{http://www.arXiv.org/abs/1507.03119}{\texttt{arXiv:1507.03119}}.

\bibitem{Aad:2015jfa}
\hrefCMSnoop {}{{ATLAS Collaboration}, ``{Measurement of the charge asymmetry
  in dileptonic decays of top quark pairs in $pp$ collisions at $\sqrt{s}=7$
  TeV using the ATLAS detector}'',} \textit{ JHEP} \textbf{ 05} (2015) 061,
  \href{http://dx.doi.org/10.1007/JHEP05(2015)061}{\doi{10.1007/JHEP05(2015)061}},
\href{http://www.arXiv.org/abs/1501.07383}{\texttt{arXiv:1501.07383}}.

\bibitem{Chatrchyan:2014yta}
\hrefCMSnoop {}{{CMS Collaboration}, ``{Measurements of the $\ttbar$ charge
  asymmetry using the dilepton decay channel in pp collisions at $\sqrt{s} =$ 7
  TeV}'',} \textit{ JHEP} \textbf{ 04} (2014) 191,
  \href{http://dx.doi.org/10.1007/JHEP04(2014)191}{\doi{10.1007/JHEP04(2014)191}},
\href{http://www.arXiv.org/abs/1402.3803}{\texttt{arXiv:1402.3803}}.

\bibitem{Aad:2016ove}
\hrefCMSnoop {}{{ATLAS Collaboration}, ``{Measurements of the charge asymmetry
  in top-quark pair production in the dilepton final state at $\sqrt{s}=8$ TeV
  with the ATLAS detector}'',} \textit{ Phys. Rev. D} \textbf{ 94} (2016)
  032006,
  \href{http://dx.doi.org/10.1103/PhysRevD.94.032006}{\doi{10.1103/PhysRevD.94.032006}},
\href{http://www.arXiv.org/abs/1604.05538}{\texttt{arXiv:1604.05538}}.

\bibitem{Khachatryan:2016ysn}
\hrefCMSnoop {}{{CMS Collaboration}, ``{Measurements of $\ttbar$ charge
  asymmetry using dilepton final states in pp collisions at $\sqrt s=8$
  TeV}'',} \textit{ Phys. Lett. B} \textbf{ 760} (2016) 365,
  \href{http://dx.doi.org/10.1016/j.physletb.2016.07.006}{\doi{10.1016/j.physletb.2016.07.006}},
\href{http://www.arXiv.org/abs/1603.06221}{\texttt{arXiv:1603.06221}}.

\bibitem{Aad:2015lgx}
\hrefCMSnoop {}{{ATLAS Collaboration}, ``{Measurement of the charge asymmetry
  in highly boosted top-quark pair production in $\sqrt{s} =$ 8 TeV $pp$
  collision data collected by the ATLAS experiment}'',} \textit{ Phys. Lett. B}
  \textbf{ 756} (2016) 52,
  \href{http://dx.doi.org/10.1016/j.physletb.2016.02.055}{\doi{10.1016/j.physletb.2016.02.055}},
\href{http://www.arXiv.org/abs/1512.06092}{\texttt{arXiv:1512.06092}}.

\bibitem{Bernreuther}
\hrefCMSnoop {}{W.~Bernreuther and Z.-G. Si, ``{Top quark and leptonic charge
  asymmetries for the Tevatron and LHC}'',} \textit{ Phys. Rev. D} \textbf{ 86}
  (2012) 034026,
  \href{http://dx.doi.org/10.1103/PhysRevD.86.034026}{\doi{10.1103/PhysRevD.86.034026}},
\href{http://www.arXiv.org/abs/1205.6580}{\texttt{arXiv:1205.6580}}.

\bibitem{nnlo2}
M.~Czakon\hrefCMSnoop {}{ {et~al.}, ``{Top-pair production at the LHC through
  NNLO QCD and NLO EW}'',} (2017).
\href{http://www.arXiv.org/abs/1705.04105}{\texttt{arXiv:1705.04105}}.

\bibitem{nnlo3}
\hrefCMSnoop {}{M.~Czakon, D.~Heymes, and A.~Mitov, ``{Dynamical scales for
  multi-TeV top-pair production at the LHC}'',} \textit{ JHEP} \textbf{ 04}
  (2017) 071,
  \href{http://dx.doi.org/10.1007/JHEP04(2017)071}{\doi{10.1007/JHEP04(2017)071}},
\href{http://www.arXiv.org/abs/1606.03350}{\texttt{arXiv:1606.03350}}.

\bibitem{nnlo4}
\hrefCMSnoop {}{M.~Czakon, P.~Fiedler, D.~Heymes, and A.~Mitov, ``{NNLO QCD
  predictions for fully-differential top-quark pair production at the
  Tevatron}'',} \textit{ JHEP} \textbf{ 05} (2016) 034,
  \href{http://dx.doi.org/10.1007/JHEP05(2016)034}{\doi{10.1007/JHEP05(2016)034}},
\href{http://www.arXiv.org/abs/1601.05375}{\texttt{arXiv:1601.05375}}.

\bibitem{nnlo1}
M.~Czakon\hrefCMSnoop {}{ {et~al.}, ``{The top-quark charge asymmetry at the
  LHC and Tevatron through NNLO QCD and NLO EW}'',} (2017).
\href{http://www.arXiv.org/abs/1711.03945}{\texttt{arXiv:1711.03945}}.

\bibitem{alpgen}
M.~L. Mangano\hrefCMSnoop {}{ {et~al.}, ``{ALPGEN, a generator for hard
  multiparton processes in hadronic collisions}'',} \textit{ JHEP} \textbf{ 07}
  (2003) 001,
  \href{http://dx.doi.org/10.1088/1126-6708/2003/07/001}{\doi{10.1088/1126-6708/2003/07/001}},
\href{http://www.arXiv.org/abs/hep-ph/0206293}{\texttt{arXiv:hep-ph/0206293}}.

\bibitem{Pumplin:2002vw}
J.~Pumplin\hrefCMSnoop {}{ {et~al.}, ``{New generation of parton distributions
  with uncertainties from global QCD analysis}'',} \textit{ JHEP} \textbf{ 07}
  (2002) 012,
  \href{http://dx.doi.org/10.1088/1126-6708/2002/07/012}{\doi{10.1088/1126-6708/2002/07/012}},
\href{http://www.arXiv.org/abs/hep-ph/0201195}{\texttt{arXiv:hep-ph/0201195}}.

\bibitem{Kersevan:2004yg}
\hrefCMSnoop {}{B.~P. Kersevan and E.~Richter-W{\c{a}}s, ``{The Monte Carlo
  event generator AcerMC versions 2.0 to 3.8 with interfaces to PYTHIA 6.4,
  HERWIG 6.5 and ARIADNE 4.1}'',} \textit{ Comput. Phys. Commun.} \textbf{ 184}
  (2013) 919,
  \href{http://dx.doi.org/10.1016/j.cpc.2012.10.032}{\doi{10.1016/j.cpc.2012.10.032}},
\href{http://www.arXiv.org/abs/hep-ph/0405247}{\texttt{arXiv:hep-ph/0405247}}.

\bibitem{Frixione:2002ik}
\hrefCMSnoop {}{S.~Frixione and B.~R. Webber, ``{Matching NLO QCD computations
  and parton shower simulations}'',} \textit{ JHEP} \textbf{ 06} (2002) 029,
  \href{http://dx.doi.org/10.1088/1126-6708/2002/06/029}{\doi{10.1088/1126-6708/2002/06/029}},
\href{http://www.arXiv.org/abs/hep-ph/0204244}{\texttt{arXiv:hep-ph/0204244}}.

\bibitem{Frixione:2008yi}
S.~Frixione\hrefCMSnoop {}{ {et~al.}, ``{Single-top hadroproduction in
  association with a W boson}'',} \textit{ JHEP} \textbf{ 07} (2008) 029,
  \href{http://dx.doi.org/10.1088/1126-6708/2008/07/029}{\doi{10.1088/1126-6708/2008/07/029}},
\href{http://www.arXiv.org/abs/0805.3067}{\texttt{arXiv:0805.3067}}.

\bibitem{Frixione:2005vw}
\hrefCMSnoop {}{S.~Frixione, E.~Laenen, P.~Motylinski, and B.~R. Webber,
  ``{Single-top production in MC@NLO}'',} \textit{ JHEP} \textbf{ 03} (2006)
  092,
  \href{http://dx.doi.org/10.1088/1126-6708/2006/03/092}{\doi{10.1088/1126-6708/2006/03/092}},
\href{http://www.arXiv.org/abs/hep-ph/0512250}{\texttt{arXiv:hep-ph/0512250}}.

\bibitem{Frixione:2007nw}
\hrefCMSnoop {}{S.~Frixione, P.~Nason, and G.~Ridolfi, ``{A Positive-weight
  next-to-leading-order Monte Carlo for heavy flavour hadroproduction}'',}
  \textit{ JHEP} \textbf{ 09} (2007) 126,
  \href{http://dx.doi.org/10.1088/1126-6708/2007/09/126}{\doi{10.1088/1126-6708/2007/09/126}},
\href{http://www.arXiv.org/abs/0707.3088}{\texttt{arXiv:0707.3088}}.

\bibitem{Re:2010bp}
\hrefCMSnoop {}{E.~Re, ``{Single-top $Wt$-channel production matched with
  parton showers using the POWHEG method}'',} \textit{ Eur. Phys. J. C}
  \textbf{ 71} (2011) 1547,
  \href{http://dx.doi.org/10.1140/epjc/s10052-011-1547-z}{\doi{10.1140/epjc/s10052-011-1547-z}},
\href{http://www.arXiv.org/abs/1009.2450}{\texttt{arXiv:1009.2450}}.

\bibitem{Alioli:2009je}
\hrefCMSnoop {}{S.~Alioli, P.~Nason, C.~Oleari, and E.~Re, ``{NLO single-top
  production matched with shower in POWHEG: $s$- and $t$-channel
  contributions}'',} \textit{ JHEP} \textbf{ 09} (2009) 111,
  \href{http://dx.doi.org/10.1088/1126-6708/2009/09/111}{\doi{10.1088/1126-6708/2009/09/111}},
  \href{http://www.arXiv.org/abs/0907.4076}{\texttt{arXiv:0907.4076}}.
[Erratum: \DOI{10.1007/JHEP02(2010)011}].

\bibitem{powheg}
\hrefCMSnoop {}{S.~Alioli, P.~Nason, C.~Oleari, and E.~Re, ``{A general
  framework for implementing NLO calculations in shower Monte Carlo programs:
  the POWHEG BOX}'',} \textit{ JHEP} \textbf{ 06} (2010) 043,
  \href{http://dx.doi.org/10.1007/JHEP06(2010)043}{\doi{10.1007/JHEP06(2010)043}},
\href{http://www.arXiv.org/abs/1002.2581}{\texttt{arXiv:1002.2581}}.

\bibitem{powheg2}
\hrefCMSnoop {}{P.~Nason, ``{A new method for combining NLO QCD with shower
  Monte Carlo algorithms}'',} \textit{ JHEP} \textbf{ 11} (2004) 040,
  \href{http://dx.doi.org/10.1088/1126-6708/2004/11/040}{\doi{10.1088/1126-6708/2004/11/040}},
\href{http://www.arXiv.org/abs/hep-ph/0409146}{\texttt{arXiv:hep-ph/0409146}}.

\bibitem{powheg3}
\hrefCMSnoop {}{S.~Frixione, P.~Nason, and C.~Oleari, ``{Matching NLO QCD
  computations with parton shower simulations: the POWHEG method}'',} \textit{
  JHEP} \textbf{ 11} (2007) 070,
  \href{http://dx.doi.org/10.1088/1126-6708/2007/11/070}{\doi{10.1088/1126-6708/2007/11/070}},
\href{http://www.arXiv.org/abs/0709.2092}{\texttt{arXiv:0709.2092}}.

\bibitem{CT10}
H.-L. Lai\hrefCMSnoop {}{ {et~al.}, ``{New parton distributions for collider
  physics}'',} \textit{ Phys. Rev. D} \textbf{ 82} (2010) 074024,
  \href{http://dx.doi.org/10.1103/PhysRevD.82.074024}{\doi{10.1103/PhysRevD.82.074024}},
\href{http://www.arXiv.org/abs/1007.2241}{\texttt{arXiv:1007.2241}}.

\bibitem{Corcella:2000bw}
G.~Corcella\hrefCMSnoop {}{ {et~al.}, ``{HERWIG 6: An event generator for
  hadron emission reactions with interfering gluons (including supersymmetric
  processes)}'',} \textit{ JHEP} \textbf{ 01} (2001) 010,
  \href{http://dx.doi.org/10.1088/1126-6708/2001/01/010}{\doi{10.1088/1126-6708/2001/01/010}},
\href{http://www.arXiv.org/abs/hep-ph/0011363}{\texttt{arXiv:hep-ph/0011363}}.

\bibitem{Butterworth:1996zw}
\hrefCMSnoop {}{J.~M. Butterworth, J.~R. Forshaw, and M.~H. Seymour,
  ``{Multiparton interactions in photoproduction at HERA}'',} \textit{ Z. Phys.
  C} \textbf{ 72} (1996) 637,
  \href{http://dx.doi.org/10.1007/s002880050286}{\doi{10.1007/s002880050286}},
\href{http://www.arXiv.org/abs/hep-ph/9601371}{\texttt{arXiv:hep-ph/9601371}}.

\bibitem{ATLAS:2011gmi}
\href {https://cds.cern.ch/record/1345343}{{ATLAS Collaboration}, ``{New ATLAS
  event generator tunes to 2010 data}'',} ATLAS report ATL-PHYS-PUB-2011-008,
  ATL-COM-PHYS-2011-329, 2011.

\bibitem{pythia}
\hrefCMSnoop {}{T.~Sj{\"o}strand, S.~Mrenna, and P.~Z. Skands, ``{PYTHIA 6.4
  physics and manual}'',} \textit{ JHEP} \textbf{ 05} (2006) 026,
  \href{http://dx.doi.org/10.1088/1126-6708/2006/05/026}{\doi{10.1088/1126-6708/2006/05/026}},
\href{http://www.arXiv.org/abs/hep-ph/0603175}{\texttt{arXiv:hep-ph/0603175}}.

\bibitem{Skands:2010ak}
\hrefCMSnoop {}{P.~Z. Skands, ``{Tuning Monte Carlo generators: The Perugia
  tunes}'',} \textit{ Phys. Rev. D} \textbf{ 82} (2010) 074018,
  \href{http://dx.doi.org/10.1103/PhysRevD.82.074018}{\doi{10.1103/PhysRevD.82.074018}},
\href{http://www.arXiv.org/abs/1005.3457}{\texttt{arXiv:1005.3457}}.

\bibitem{MM}
\hrefCMSnoop {}{{ATLAS Collaboration}, ``{Measurement of the top quark-pair
  production cross section with ATLAS in $pp$ collisions at $\sqrt{s}=7$
  TeV}'',} \textit{ Eur. Phys. J. C} \textbf{ 71} (2011) 1577,
  \href{http://dx.doi.org/10.1140/epjc/s10052-011-1577-6}{\doi{10.1140/epjc/s10052-011-1577-6}},
\href{http://www.arXiv.org/abs/1012.1792}{\texttt{arXiv:1012.1792}}.

\bibitem{btag1}
\href {http://cds.cern.ch/record/1664335}{{ATLAS Collaboration}, ``{Calibration
  of $b$-tagging using dileptonic top pair events in a combinatorial likelihood
  approach with the ATLAS experiment}'',} {ATLAS} note {ATLAS-CONF-2014-004},
  2014.

\bibitem{btag2}
\hrefCMSnoop {}{{ATLAS Collaboration}, ``{Calibration of the performance of
  $b$-tagging for $c$ and light-flavour jets in the 2012 ATLAS data}'',}
  \textit{ ATLAS-CONF-2014-046}
(2014).

\bibitem{klfitter}
J.~Erdmann\hrefCMSnoop {}{ {et~al.}, ``{A likelihood-based reconstruction
  algorithm for top-quark pairs and the KLFitter framework}'',} \textit{ Nucl.
  Instrum. Meth. A} \textbf{ 748} (2014) 18,
  \href{http://dx.doi.org/10.1016/j.nima.2014.02.029}{\doi{10.1016/j.nima.2014.02.029}},
\href{http://www.arXiv.org/abs/1312.5595}{\texttt{arXiv:1312.5595}}.

\bibitem{fbu}
\hrefCMSnoop {}{G.~Choudalakis, ``{Fully Bayesian Unfolding}'',} (2012).
\href{http://www.arXiv.org/abs/1201.4612}{\texttt{arXiv:1201.4612}}.

\bibitem{madgraph}
J.~Alwall\hrefCMSnoop {}{ {et~al.}, ``{MadGraph 5: going beyond}'',} \textit{
  JHEP} \textbf{ 06} (2011) 128,
  \href{http://dx.doi.org/10.1007/JHEP06(2011)128}{\doi{10.1007/JHEP06(2011)128}},
\href{http://www.arXiv.org/abs/1106.0522}{\texttt{arXiv:1106.0522}}.

\bibitem{Chatrchyan:2011id}
\hrefCMSnoop {}{{CMS Collaboration}, ``{Measurement of the underlying event
  activity at the LHC with $\sqrt{s}= 7$ TeV and comparison with $\sqrt{s} =
  0.9$ TeV}'',} \textit{ JHEP} \textbf{ 09} (2011) 109,
  \href{http://dx.doi.org/10.1007/JHEP09(2011)109}{\doi{10.1007/JHEP09(2011)109}},
\href{http://www.arXiv.org/abs/1107.0330}{\texttt{arXiv:1107.0330}}.

\bibitem{Khachatryan:2015pea}
\hrefCMSnoop {}{{CMS Collaboration}, ``{Event generator tunes obtained from
  underlying event and multiparton scattering measurements}'',} \textit{ Eur.
  Phys. J. C} \textbf{ 76} (2016) 155,
  \href{http://dx.doi.org/10.1140/epjc/s10052-016-3988-x}{\doi{10.1140/epjc/s10052-016-3988-x}},
\href{http://www.arXiv.org/abs/1512.00815}{\texttt{arXiv:1512.00815}}.

\bibitem{Chatrchyan:2013gfi}
\hrefCMSnoop {}{{CMS Collaboration}, ``{Study of the underlying event at
  forward rapidity in pp collisions at $\sqrt{s}$ = 0.9, 2.76, and 7 TeV}'',}
  \textit{ JHEP} \textbf{ 04} (2013) 072,
  \href{http://dx.doi.org/10.1007/JHEP04(2013)072}{\doi{10.1007/JHEP04(2013)072}},
\href{http://www.arXiv.org/abs/1302.2394}{\texttt{arXiv:1302.2394}}.

\bibitem{csvtagger}
\hrefCMSnoop {}{{CMS Collaboration}, ``{Identification of $b$-quark jets with
  the CMS experiment}'',} \textit{ JINST} \textbf{ 8} (2013) P04013,
  \href{http://dx.doi.org/10.1088/1748-0221/8/04/P04013}{\doi{10.1088/1748-0221/8/04/P04013}},
\href{http://www.arXiv.org/abs/1211.4462}{\texttt{arXiv:1211.4462}}.

\bibitem{TUnfold}
\hrefCMSnoop {}{S.~Schmitt, ``{TUnfold: an algorithm for correcting migration
  effects in high energy physics}'',} \textit{ JINST} \textbf{ 7} (2012)
  T10003,
  \href{http://dx.doi.org/10.1088/1748-0221/7/10/T10003}{\doi{10.1088/1748-0221/7/10/T10003}},
\href{http://www.arXiv.org/abs/1205.6201}{\texttt{arXiv:1205.6201}}.

\bibitem{blue1}
\hrefCMSnoop {}{L.~Lyons, D.~Gibaut, and P.~Clifford, ``How to combine
  correlated estimates of a single physical quantity'',} \textit{ Nucl.
  Instrum. Meth. A} \textbf{ 270} (1988) 110,
\href{http://dx.doi.org/10.1016/0168-9002(88)90018-6}{\doi{10.1016/0168-9002(88)90018-6}}.

\bibitem{blue2}
\hrefCMSnoop {}{A.~Valassi, ``{Combining correlated measurements of several
  different physical quantities}'',} \textit{ Nucl. Instrum. Meth. A} \textbf{
  500} (2003) 391,
\href{http://dx.doi.org/10.1016/S0168-9002(03)00329-2}{\doi{10.1016/S0168-9002(03)00329-2}}.

\bibitem{blue_implementation}
\hrefCMSnoop {}{R.~Nisius, ``{On the combination of correlated estimates of a
  physics observable}'',} \textit{ Eur. Phys. J. C} \textbf{ 74} (2014) 3004,
  \href{http://dx.doi.org/10.1140/epjc/s10052-014-3004-2}{\doi{10.1140/epjc/s10052-014-3004-2}},
\href{http://www.arXiv.org/abs/1402.4016}{\texttt{arXiv:1402.4016}}.

\bibitem{JESATLASCMS7TeV}
\hrefCMSnoop {}{{ATLAS and CMS Collaborations}, ``{Jet energy scale uncertainty
  correlations between ATLAS and CMS}'',} \textit{ ATL-PHYS-PUB-2014-020,
  CMS-PAS-JME-14-003} (2014).

\bibitem{CTEQ6.6}
P.~M. Nadolsky\hrefCMSnoop {}{ {et~al.}, ``{Implications of CTEQ global
  analysis for collider observables}'',} \textit{ Phys. Rev. D} \textbf{ 78}
  (2008) 013004,
  \href{http://dx.doi.org/10.1103/PhysRevD.78.013004}{\doi{10.1103/PhysRevD.78.013004}},
\href{http://www.arXiv.org/abs/0802.0007}{\texttt{arXiv:0802.0007}}.

\bibitem{MSTW08}
\hrefCMSnoop {}{A.~D. Martin, W.~J. Stirling, R.~S. Thorne, and G.~Watt,
  ``{Parton distributions for the LHC}'',} \textit{ Eur. Phys. J. C} \textbf{
  63} (2009) 189,
  \href{http://dx.doi.org/10.1140/epjc/s10052-009-1072-5}{\doi{10.1140/epjc/s10052-009-1072-5}},
\href{http://www.arXiv.org/abs/0901.0002}{\texttt{arXiv:0901.0002}}.

\bibitem{NNPDF2.1}
R.~D. Ball\hrefCMSnoop {}{ {et~al.}, ``{Impact of heavy quark masses on parton
  distributions and LHC phenomenology}'',} \textit{ Nucl. Phys. B} \textbf{
  849} (2011) 296,
  \href{http://dx.doi.org/10.1016/j.nuclphysb.2011.03.021}{\doi{10.1016/j.nuclphysb.2011.03.021}},
\href{http://www.arXiv.org/abs/1101.1300}{\texttt{arXiv:1101.1300}}.

\bibitem{PDF4LHC}
\hrefCMSnoop {}{M.~Botje {et~al.}, ``{The PDF4LHC Working Group Interim
  Recommendations}'',} (2011).
\href{http://www.arXiv.org/abs/1101.0538}{\texttt{arXiv:1101.0538}}.

\bibitem{Cacciari:2008gp}
\hrefCMSnoop {}{M.~Cacciari, G.~P. Salam, and G.~Soyez, ``The anti-$k_t$ jet
  clustering algorithm'',} \textit{ JHEP} \textbf{ 04} (2008) 063,
  \href{http://dx.doi.org/10.1088/1126-6708/2008/04/063}{\doi{10.1088/1126-6708/2008/04/063}},
  \href{http://www.arXiv.org/abs/0802.1189}{\texttt{arXiv:0802.1189}}.

\bibitem{PF}
\hrefCMSnoop {}{{CMS Collaboration}, ``Particle-flow reconstruction and global
  event description with the {CMS} detector'',} \textit{ JINST} \textbf{ 12}
  (2017) P10003,
  \href{http://dx.doi.org/10.1088/1748-0221/12/10/P10003}{\doi{10.1088/1748-0221/12/10/P10003}},
\href{http://www.arXiv.org/abs/1706.04965}{\texttt{arXiv:1706.04965}}.

\bibitem{JESATLASCMS}
\href {http://cds.cern.ch/record/2104039}{{ATLAS and CMS Collaborations},
  ``{Jet energy scale uncertainty correlations between ATLAS and CMS at
  $8\,\mathrm{TeV}$}'',} CMS Physics Analysis Summary/ATLAS Note
  ATL-PHYS-PUB-2015-049, CMS-PAS-JME-15-001, 2015.

\bibitem{Khachatryan:2015oqa}
\hrefCMSnoop {}{{CMS Collaboration}, ``{Measurement of the differential cross
  section for top quark pair production in $pp$ collisions at $\sqrt{s} =
  8\,\text {TeV} $}'',} \textit{ Eur. Phys. J. C} \textbf{ 75} (2015) 542,
  \href{http://dx.doi.org/10.1140/epjc/s10052-015-3709-x}{\doi{10.1140/epjc/s10052-015-3709-x}},
\href{http://www.arXiv.org/abs/1505.04480}{\texttt{arXiv:1505.04480}}.

\bibitem{ATLAS_diffxsec8TeV}
\hrefCMSnoop {}{{ATLAS Collaboration}, ``{Measurements of top-quark pair
  differential cross-sections in the lepton+jets channel in $pp$ collisions at
  $\sqrt{s}=8$ TeV using the ATLAS detector}'',} \textit{ Eur. Phys. J. C}
  \textbf{ 76} (2016) 538,
  \href{http://dx.doi.org/10.1140/epjc/s10052-016-4366-4}{\doi{10.1140/epjc/s10052-016-4366-4}},
\href{http://www.arXiv.org/abs/1511.04716}{\texttt{arXiv:1511.04716}}.

\bibitem{AguilarSaavedra:2011hz}
\hrefCMSnoop {}{J.~A. Aguilar-Saavedra and M.~Perez-Victoria, ``{Asymmetries in
  $t\bar{t}$ production: LHC versus Tevatron}'',} \textit{ Phys. Rev. D}
  \textbf{ 84} (2011) 115013,
  \href{http://dx.doi.org/10.1103/PhysRevD.84.115013}{\doi{10.1103/PhysRevD.84.115013}},
\href{http://www.arXiv.org/abs/1105.4606}{\texttt{arXiv:1105.4606}}.

\bibitem{AguilarSaavedra:2011ug}
\hrefCMSnoop {}{J.~A. Aguilar-Saavedra and M.~Perez-Victoria, ``{Simple models
  for the top asymmetry: Constraints and predictions}'',} \textit{ JHEP}
  \textbf{ 09} (2011) 097,
  \href{http://dx.doi.org/10.1007/JHEP09(2011)097}{\doi{10.1007/JHEP09(2011)097}},
\href{http://www.arXiv.org/abs/1107.0841}{\texttt{arXiv:1107.0841}}.

\bibitem{Aguilar-Saavedra:2014nja}
\hrefCMSnoop {}{J.~A. Aguilar-Saavedra, ``{Portrait of a colour octet}'',}
  \textit{ JHEP} \textbf{ 08} (2014) 172,
  \href{http://dx.doi.org/10.1007/JHEP08(2014)172}{\doi{10.1007/JHEP08(2014)172}},
\href{http://www.arXiv.org/abs/1405.5826}{\texttt{arXiv:1405.5826}}.

\end{thebibliography}\endgroup

\cleardoublepage \appendix
\section{The CMS Collaboration \label{app:collab}}
\begin{sloppypar}\hyphenpenalty=5000\widowpenalty=500\clubpenalty=5000
\vskip\cmsinstskip
\textbf{Yerevan Physics Institute,  Yerevan,  Armenia}\\*[0pt]
A.M.~Sirunyan,  A.~Tumasyan
\vskip\cmsinstskip
\textbf{Institut f\"{u}r Hochenergiephysik,  Wien,  Austria}\\*[0pt]
W.~Adam,  F.~Ambrogi,  E.~Asilar,  T.~Bergauer,  J.~Brandstetter,  E.~Brondolin,  M.~Dragicevic,  J.~Er\"{o},  A.~Escalante Del Valle,  M.~Flechl,  M.~Friedl,  R.~Fr\"{u}hwirth\cmsAuthorMark{1},  V.M.~Ghete,  J.~Grossmann,  J.~Hrubec,  M.~Jeitler\cmsAuthorMark{1},  A.~K\"{o}nig,  N.~Krammer,  I.~Kr\"{a}tschmer,  D.~Liko,  T.~Madlener,  I.~Mikulec,  E.~Pree,  N.~Rad,  H.~Rohringer,  J.~Schieck\cmsAuthorMark{1},  R.~Sch\"{o}fbeck,  M.~Spanring,  D.~Spitzbart,  A.~Taurok,  W.~Waltenberger,  J.~Wittmann,  C.-E.~Wulz\cmsAuthorMark{1},  M.~Zarucki
% ...
\vskip\cmsinstskip
% ...
1:~Also at Vienna University of Technology,  Vienna,  Austria\\
% ...
\end{sloppypar}
\section{The ATLAS Collaboration \label{app:ATLAScollab}} \begin{sloppypar}\hyphenpenalty=5000\widowpenalty=500\clubpenalty=5000

{\Large The ATLAS Collaboration}
\begin{flushleft}
\bigskip

M.~Aaboud$^\textrm{\scriptsize 137d}$,
G.~Aad$^\textrm{\scriptsize 88}$,
B.~Abbott$^\textrm{\scriptsize 115}$,
O.~Abdinov$^\textrm{\scriptsize 12}$$^{,*}$,
B.~Abeloos$^\textrm{\scriptsize 119}$,
S.H.~Abidi$^\textrm{\scriptsize 161}$,
O.S.~AbouZeid$^\textrm{\scriptsize 139}$,
N.L.~Abraham$^\textrm{\scriptsize 151}$,
H.~Abramowicz$^\textrm{\scriptsize 155}$,
H.~Abreu$^\textrm{\scriptsize 154}$
% ...
\bigskip
\\
% ...
$^{12}$ Institute of Physics, Azerbaijan Academy of Sciences, Baku, Azerbaijan\\
% ...
$^{88}$ CPPM, Aix-Marseille Universit{\'e} and CNRS/IN2P3, Marseille, France\\
% ...
$^{115}$ Homer L. Dodge Department of Physics and Astronomy, University of Oklahoma, Norman OK, United States of America\\
% ...
$^{119}$ LAL, Univ. Paris-Sud, CNRS/IN2P3, Universit{\'e} Paris-Saclay, Orsay, France\\
% ...
$^{137}$ $^{(a)}$ Facult{\'e} des Sciences Ain Chock, R{\'e}seau Universitaire de Physique des Hautes Energies - Universit{\'e} Hassan II, Casablanca; $^{(b)}$ Centre National de l'Energie des Sciences Techniques Nucleaires, Rabat; $^{(c)}$ Facult{\'e} des Sciences Semlalia, Universit{\'e} Cadi Ayyad, LPHEA-Marrakech; $^{(d)}$ Facult{\'e} des Sciences, Universit{\'e} Mohamed Premier and LPTPM, Oujda; $^{(e)}$ Facult{\'e} des sciences, Universit{\'e} Mohammed V, Rabat, Morocco\\
% ...
$^{139}$ Santa Cruz Institute for Particle Physics, University of California Santa Cruz, Santa Cruz CA, United States of America\\
% ...
$^{151}$ Department of Physics and Astronomy, University of Sussex, Brighton, United Kingdom\\
% ...
$^{154}$ Department of Physics, Technion: Israel Institute of Technology, Haifa, Israel\\
$^{155}$ Raymond and Beverly Sackler School of Physics and Astronomy, Tel Aviv University, Tel Aviv, Israel\\
% ...
$^{161}$ Department of Physics, University of Toronto, Toronto ON, Canada\\
% ...
$^{a}$ Also at Department of Physics, King's College London, London, United Kingdom\\
$^{b}$ Also at Institute of Physics, Azerbaijan Academy of Sciences, Baku, Azerbaijan\\
$^{c}$ Also at Novosibirsk State University, Novosibirsk, Russia\\
$^{d}$ Also at TRIUMF, Vancouver BC, Canada\\
$^{e}$ Also at Department of Physics {\&} Astronomy, University of Louisville, Louisville, KY, United States of America\\
% ...
$^{*}$ Deceased
\end{flushleft}

\end{sloppypar}
\end{document}